\documentclass[aps,prb,twocolumn,groupedaddress]{revtex4}

\usepackage{graphicx}

\begin{document}

\title{A model metal potential exhibiting polytetrahedral clusters}

\author{Jonathan P.~K.~Doye}
\email[]{jpkd1@cam.ac.uk}
\affiliation{University Chemical Laboratory, Lensfield Road, Cambridge CB2 1EW, United Kingdom} 

\date{\today}

\begin{abstract}
Putative global minima have been located for clusters
interacting with an aluminium glue potential for $N\le 190$.
Virtually all the clusters have polytetrahedral structures,
which for larger sizes involve an ordered array of disclinations
that are similar to those in 
the Z, H and $\sigma$ Frank-Kasper phases.
Comparisons of sequences of larger clusters suggest that the majority
of the global minima will adopt the bulk face-centred-cubic
 structure beyond $N\approx 500$.
\end{abstract}

\maketitle

\section{\label{sect:intro}Introduction}
The understanding of the structure of metal clusters has seen many developments
in recent years,\cite{Alonso00,Johnston02} however there is still much to be 
learnt. 
From the theoretical perspective, for example, only relatively recently
have global optimization techniques become sufficiently powerful to find
the most stable structures of metal clusters with up to 100 
atoms,\cite{WalesS99} even
when described by relatively simple many-body potentials. 
This has led to many interesting new structures being 
revealed\cite{Garzon98,Doye98c,Wilson00,Soler01,Wang01,Michaelian02,Lai02,Doye03a}
that go beyond those classified for pair potentials, but 
there are still probably many classes of structure that remain undiscovered. 
Furthermore, the many-body nature of the bonding makes it difficult
to unravel the relationship between the observed structures and the
form of the interactions,
although there has been some interesting recent 
progress.\cite{Soler00,Baletto02b,Doye03a} 
This situation contrasts with that for 
pair potentials, where, for example, the effects of the width of the 
potential well\cite{Doye95c,Doye97d}
and oscillations in the potential\cite{Doye01a,Doye01d,Doye03b} 
on the cluster structure have been mapped out and rationalized.

One of the more interesting recent discoveries has been
that some monatomic metal clusters\cite{Dassenoy00,Wang01}  
can exhibit structures with polytetrahedral order,\cite{NelsonS,Sadoc99}
i.e. the whole of the cluster can be naturally divided up 
into tetrahedra with atoms at their vertices.\cite{closep}
In the experiments of Dassenoy {\it et al.}\ on cobalt clusters,
the electron microscopy images and diffraction patterns 
could not be explained by any of the usual structural
forms for a cluster, but instead were consistent with a 
polytetrahedral structure.\cite{Dassenoy00} 
Furthermore, the magic number at 
$N=61$ in the mass spectrum of strontium
clusters\cite{Brechignac00,Wang01} was 
assigned to a polytetrahedral structure,\cite{Wang01} 
and the next magic number at $N=82$ also coincides with
a polytetrahedral magic number.\cite{Doye01d}
Similar magic number patterns have
also been found for some rare-earth metals.\cite{Brechignac93}

\begin{figure}
\includegraphics[width=8.4cm]{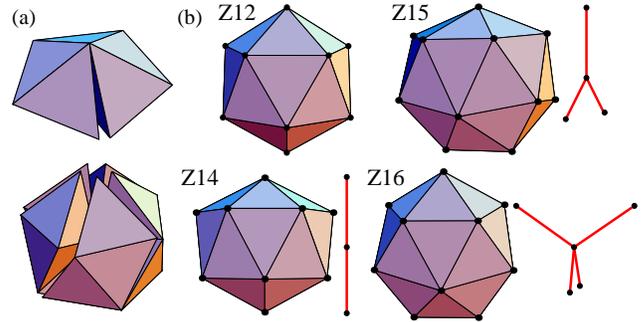}
\caption{\label{fig:gaps}(a) Packings of regular 
tetrahedra to illustrate the strain inherent in polytetrahedral packings. 
Five regular tetrahedra around a common edge produce a gap of $7.36^\circ$, and 
twenty regular tetrahedra about a common vertex
produce gaps equivalent to a solid angle of 1.54 steradians.
(b) The Frank-Kasper polyhedra for coordination numbers 12, 14, 15 and 16, 
as labelled.  The associated disclination networks are
placed next to each cluster.}
\end{figure}

These results seem somewhat surprising since 
ordered polytetrahedral structures are restricted to alloys in bulk, 
e.g the crystalline Frank-Kasper phases\cite{FrankK58,FrankK59,Shoemaker,binary}
and quasicrystals,\cite{Shechtman84} structural models of which often invoke 
substantial polytetrahedral order.\cite{Elser85,Sachdev85b,Audier86}
However, as we will see, there are good reasons why polytetrahedral 
structures are more likely to be seen for monatomic systems in clusters.

Figure \ref{fig:gaps}(a) illustrates that all space cannot be packed
with regular tetrahedra. (For this reason polytetrahedral structures 
are sometimes said to be frustrated.\cite{Sadoc99}) 
Therefore, the nearest-neighbour pair distances in a polytetrahedral 
structure cannot all take the same value. 
For example to close the gaps associated with the icosahedral
packing of regular tetrahedra, the tetrahedra must be distorted such that
the edges of the tetrahedra 
on the surface are 5\% longer than the radial edges.
For a monatomic system this strain usually leads to an energetic
penalty, although for a binary system, if a smaller atom of the right size 
is placed in the centre of the icosahedron, all pair distances could still
take their ideal values.

For clusters of sufficiently small size the 
energetic penalties associated with the strain 
inherent to polytetrahedral structures
can often be more than compensated by a favourable surface energy.
The surface of a polytetrahedral cluster is 
triangulated leading to high coordination numbers, whereas
clusters based on a close-packing scheme may 
have to expose $\{100\}$ faces, which involve atoms with lower
coordination number, and hence typically have a higher surface energy.

The 13-atom icosahedral structure,
which is very common for systems interacting with both pair potentials 
and many-body metal potentials, illustrates this principle. 
The polytetrahedral packing of the icosahedron can be continued 
by adding capping atoms above the faces and vertices.\cite{Northby87} 
This ``anti-Mackay'' overlayer leads to polyicosahedral 
structures, where each atom in the interior of the cluster has 
local icosahedral coordination. 
So, the 19-atom double icosahedron is also common, but as
the size increases these structures become
less likely to be observed. For example, for Lennard-Jones clusters
they are lowest in energy up to $N=30$.\cite{Northby87}  
The reason that these clusters
become disfavoured is clear from Fig.\ \ref{fig:gaps}(a). As further 
regular tetrahedra are added to these packings the gaps involved
become larger and larger, and hence the strain energy associated
with polyicosahedral packings increases very rapidly with size.

For the above polytetrahedral clusters, there are five tetrahedra
around each edge in the interior of the cluster.
To balance out the increasing gaps in this type of packing, the 
bulk Frank-Kasper phases also involve sites where six tetrahedra
share a common edge---a negative disclination is said to run
along this edge.\cite{disclin} Although the local distortion required to remove 
the overlap when six regular tetrahedra are placed around a common
edge  is larger, the introduction of disclination lines leads 
to a decrease in the overall strain.

The coordination polyhedra involved in the 
Frank-Kasper phases\cite{FrankK58,FrankK59} are
depicted in Figure \ref{fig:gaps}(b). Local icosahedral coordination
involves no disclinations, but as the coordination number increases
negative disclinations must be introduced.
Note a disclination line can only end by forming a loop or 
exiting at a surface.
So for the Z14 coordination polyhedron the disclination passes through
the central atom, 
and for the Z15 and Z16 coordination polyhedra the 
central atom acts as a node for three and four disclinations, respectively.
Polytetrahedral packings can therefore be described by a network 
of disclination lines threading an icosahedrally-coordinated medium.
In Frank-Kasper crystals this disclination network is ordered 
and periodic,\cite{major} and is mediated by the Z14, Z15 and Z16 
Frank-Kasper coordination polyhedra. 

Polytetrahedral clusters that involve disclinations have previously been 
found for two very different pair potentials. Firstly, for clusters interacting
with a sufficiently long-ranged Morse potential the strain 
involved in a polytetrahedral structure can be tolerated, 
and ordered polytetrahedral clusters are found up to 
$N\approx 70$.\cite{Doye97d}
Secondly, a modified Dzugutov potential favours clusters that are based 
on the C14 and C15 Frank-Kasper phases up to at least $N=250$.\cite{Doye01d} 
This potential has a maximum in the potential
that has been chosen to coincide with the 
next-nearest neighbour distance in close-packed structures,
and somewhat resembles the Friedel oscillations that can occur 
in effective pair potentials derived for metallic 
systems.\cite{Hafner88b,Moriarty97}

These two examples are for somewhat unusual potentials.
However, polytetrahedral structures are perhaps more likely for metals,
because the nature of the metallic potentials makes them potentially
less sensitive to the strain in the individual pair distances,\cite{Soler00}
for reasons that will be discussed in more detail in 
Section \ref{sect:potential}.
For example, similar structures to those of the Morse clusters have also 
been found for clusters described by a generalized set of 
metallic interactions.\cite{Cune00}
Furthermore, in this paper I report that the lowest-energy structures 
of clusters interacting with an aluminium glue potential\cite{Ercolessi94}  
are also polytetrahedral up to at least $N=190$. 

The paper is organized as follows. In Section \ref{sect:methods} the 
main focus is on the form of the aluminium potential and the implications
of this form for cluster structure. In Section \ref{sect:gmin} I 
describe in detail the structures of the putative global minima,
and in Section \ref{sect:Al61} I take Al$_{61}$ as a case
study for understanding the reasons for this system's preference
for polytetrahedral structures.
Then, in Section \ref{sect:large} I briefly examine the size evolution
of the cluster structure at larger sizes.
The emphasis in the paper is less on the clusters as a realistic
model of aluminium clusters that will hold up to detailed scrutiny---this 
is very difficult for any empirical potential---and more on using
these clusters as a model system for exploring the types of polytetrahedral
structure that metal clusters might exhibit, and the reasons for their
formation.
However, in Section \ref{sect:large} I do compare the results
with the available experimental information on the structure of 
aluminium clusters.

\section{\label{sect:methods}Methods}
\subsection{\label{sect:potential}Potential}

To model the aluminium clusters I use a glue 
potential\cite{Ercolessi88} that has been 
constructed using the force-matching method.\cite{Ercolessi94} 
The potential energy is given by
\begin{eqnarray}
E&=&E_{\rm pair} + E_{\rm glue} \\
 &=&\sum_{i<j} \phi\left(r_{ij}\right)+\sum_{i} U\left(\bar\rho_i\right),
\end{eqnarray}
where $\phi(r)$ is a short-ranged pair potential, $U(\bar\rho)$ is a many-body glue function
and $\bar\rho_i$ is defined as 
\begin{equation}
\bar\rho_i=\sum_j \rho\left(r_{ij}\right),
\end{equation}
where $\rho(r)$ is an ``atomic density'' function.
These three functions have been fitted to match the forces
produced by first-principles electronic structure calculations for a large set of configurations,
including those that correspond to surfaces, clusters, liquids and crystals.\cite{Ercolessi94} 
That the potential has been designed to model a wide variety of atomic
environments, such as occur in clusters, 
is particularly important for the current application,
as opposed to a potential that has just been fitted to bulk data.
This flexibility has also led the potential to be used in a wide
variety of applications, often with considerable success.
Examples include clusters,\cite{Sun98} wires,\cite{Gulseren98} 
surface thermodynamics,\cite{DiTolla94,Raphuthi95} 
grain boundaries,\cite{Rittner96} 
bulk\cite{Sandberg02}  and surface\cite{Trushin01} diffusion and
melt solidification.\cite{Liu01}

\begin{figure}
\includegraphics[width=8.4cm]{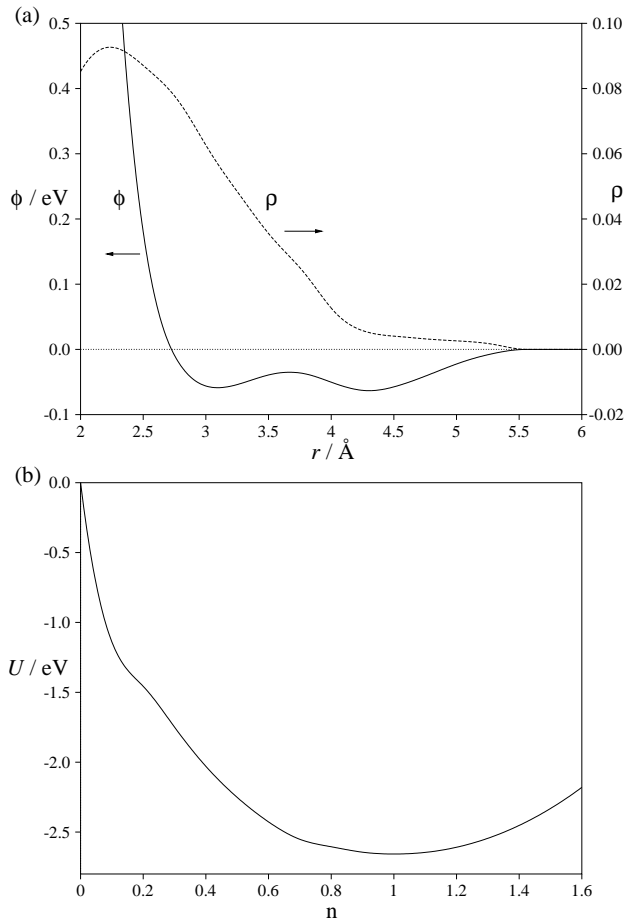}
\caption{\label{fig:potential}The three functions that make up the aluminium glue potential:
(a) $\phi(r)$, $\rho(r)$ and (b) $U(\bar\rho)$.}
\end{figure}

The functions $\phi(r)$, $\rho(r)$ and $U(\bar\rho)$ are displayed in 
Figure \ref{fig:potential}. 
The forms of these functions are not assumed, but are determined
by the fitting process. This contrasts with other empirical 
potentials, such as the Gupta or Sutton-Chen potentials. With
these latter types of potentials it would be impossible, for example, to
reproduce the double well structure of $\phi(r)$, which 
also occurs for a sophisticated `bulk' potential.\cite{Mishin99}

The shape of these functions have some fairly straightforward
consequences for the cluster structure. 
$U(\bar\rho)$ has a single minimum of depth $-2.6571\,$eV at $\bar\rho=1.0$.
This term provides the dominant contribution to the energy, 
so it is important that each atom has a value of $\bar\rho$ as close 
as possible to one. Of course,
the values associated with the surface atoms will be significantly less than this.
There is therefore a tendency for surface distances to contract so
that the atoms achieve as high a value of $\bar\rho$ as is possible. 
This effect will be balanced by
the energetic penalty associated with the resulting compression of the 
interior and the reduction in the pair energy of the surface atoms. 
Furthermore, the minimum of the glue function is rather broad, 
and so the energy loss associated with $\bar\rho$ values that deviate 
from one is not so severe. Thus, surface contraction will not be 
as pronounced as for some other metallic systems. For example, 
the lead glue potential,\cite{Lim} whose cluster structures have 
recently been analysed,\cite{Hendy02,Doye03a}
has a stronger dependence of $U$ on $\bar\rho$.

$\rho(r)$ decreases roughly linearly with distance
until $r\approx 4\,$\AA, after which it decays more slowly to zero 
at the cutoff distance.
As the value of $\rho(r)$ in the latter regime is always small, 
the major contribution to the $\bar\rho$ values of each atom comes 
from nearest neighbours, and hence the average value of $\bar\rho_i$ will 
be maximized by clusters with a large number of nearest neighbours.  

$\phi(r)$ has an interesting double-well structure. 
The first minimum occurs at $r_{\rm min1}=3.095\,{\rm\AA}$, 
and has a depth of $-0.0587\,$eV. 
This is separated by a maximum at $3.669\,{\rm\AA}$ from a 
second minimum at $r_{\rm min2}=4.303\,{\rm\AA}=1.390\, r_{\rm min1}$
of slightly greater depth, $-0.0633\,$eV.
As the number of next-nearest neighbours will be significantly larger 
than nearest neighbours (except for the smallest clusters), the largest
contribution to $E_{\rm pair}$ comes from the second neighbour shell.

The ratio of the positions of the two minima would initially suggest that they
are almost ideally placed to coincide with the first and second shells for 
closed-packed structures, which have a ratio of $\sqrt 2$.
However, the nearest-neighbour distance is determined more 
by the glue part of the potential. 
For example, for an atom with twelve nearest neighbours, 
these neighbours would need to be at a distance 
for 2.611\,\AA\ to achieve a $\bar\rho$ value of 1.
If the effects of next neighbours are also included, this distance does not 
not need to be as short, and so 
the nearest neighbour distance for the equilibrium face-centred-cubic (fcc) 
crystal, $r_{\rm nn}^{\rm xtal}$, is 2.851\,\AA.\cite{Ercolessi94}
Compared to this distance the position of the second minimum is not so 
advantageous for closed-packed structures; 
$r_{\rm min2}=1.509\, r_{\rm nn}^{\rm xtal}$.

One of the important features of the glue energy for the current study 
is that it does not directly depend on the distribution of pair distances 
or the coordination number of an atom, but only on the $\bar\rho_i$ 
values.\cite{Soler00} 
This situation contrasts with that for a pair potential, and makes 
metallic systems potentially less sensitive to internal strains. 
This feature has already been cited as an important factor in the
stability of disordered clusters for a variety of metallic 
systems.\cite{Soler00} The same principle applies for ordered polytetrahedral 
clusters, which have both strain and high coordination numbers.
For example, the ideal glue energy can be achieved
for an atom with coordination number higher than twelve simply by 
expanding the nearest-neighbour distances for this atom with respect 
to $r_{\rm nn}^{\rm xtal}$ until $\bar\rho=1$ for that atom.
Similarly, there is no reason why an atom with a wide distribution of 
nearest-neighbour distances cannot also achieve a $\bar\rho$ value 
of one.

\subsection{\label{sect:gopt}Global Optimization}

\begin{figure*}
\includegraphics[width=18cm]{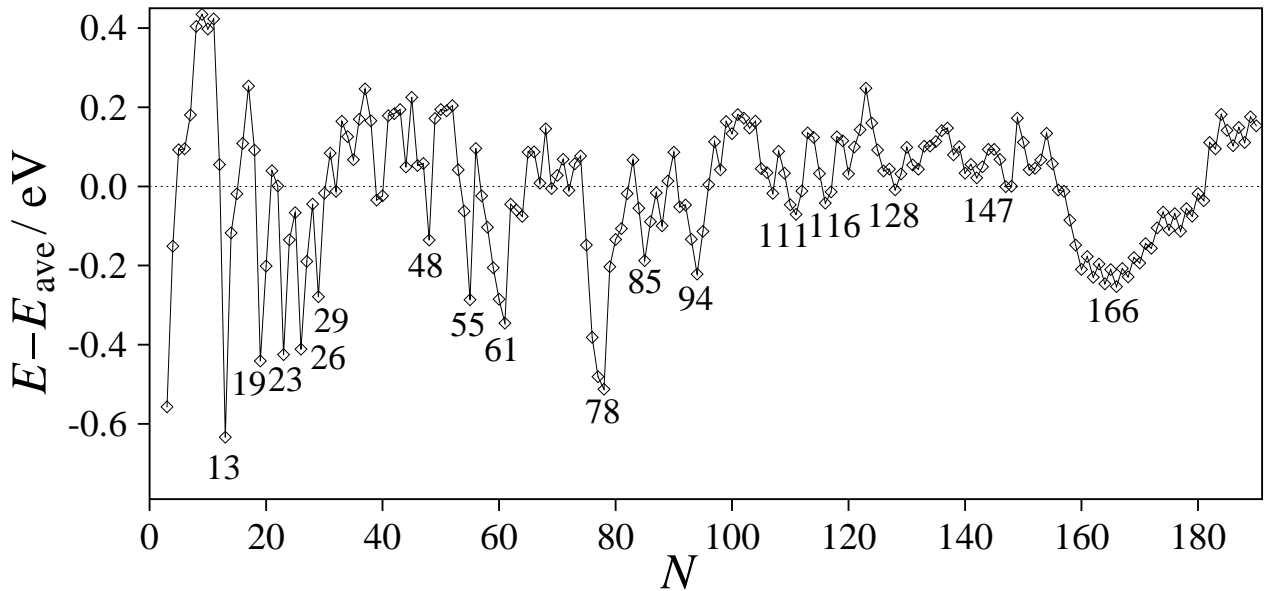}
\caption{\label{fig:EvN}Energies of the putative global minima relative to 
$E_{\rm ave}$, a four-parameter fit to their energies.
$E_{\rm ave}=-3.352 N+2.001 N^{2/3} - 0.531 N^{1/3} + 3.119$.}
\end{figure*}

The global optimization of the aluminium clusters 
was performed using the basin-hopping \cite{WalesD97,WalesS99}
(or Monte Carlo minimization \cite{Li87a}) approach. 
This method has proved particularly
successful in locating putative global minima for a 
wide variety of cluster systems.
As the results are extended up to 190 atoms, a considerable
computational effort was required. 
It proved particularly important to supplement the
standard unbiased runs from random starting points for each size, 
with runs started from low-energy minima of nearby sizes with the 
appropriate number of atoms added or removed. These latter runs were
applied iteratively until no further new global minima were located.

It should be noted that, of course, 
there is no guarantee that I have been able to locate the true global 
minima, and the probability that a global minimum has been missed
will increase with cluster size, 
as the size of the search space, and hence the number of 
minima,\cite{Tsai93a,Still99,Doye02a} increases exponentially with $N$.
However, from examination of the statistics of how often independent runs locate the 
same lowest-energy minimum, I am confident that virtually all of the 
putative global minima up to $N=100$ cannot be bettered, and
that beyond this size the putative global minima, if not truly global,
are very good estimates. In this latter size range the runs seeded from
low-energy clusters of other sizes become more important. 
Therefore, global minima are most likely to be missed when they are 
structurally different from those of nearby sizes.
For example, the lowest-energy structure found in a basin-hopping run 
for Al$_{147}$ was at $-436.6892\,$eV, even though the 147-atom 
Mackay icosahedron has an energy that is $0.0133\,$eV lower.
However, this is the only example I found where reoptimization of a structure
that is known to often be particularly stable bettered the global 
optimization results. I also tried to construct large polytetrahedral 
structures based upon some of the structural motifs present for $N<100$, 
but again this did not lead to any improvements.

\section{\label{sect:gmin}Global minima for $N\le 190$}

\begin{figure*}
\includegraphics[width=14.75cm]{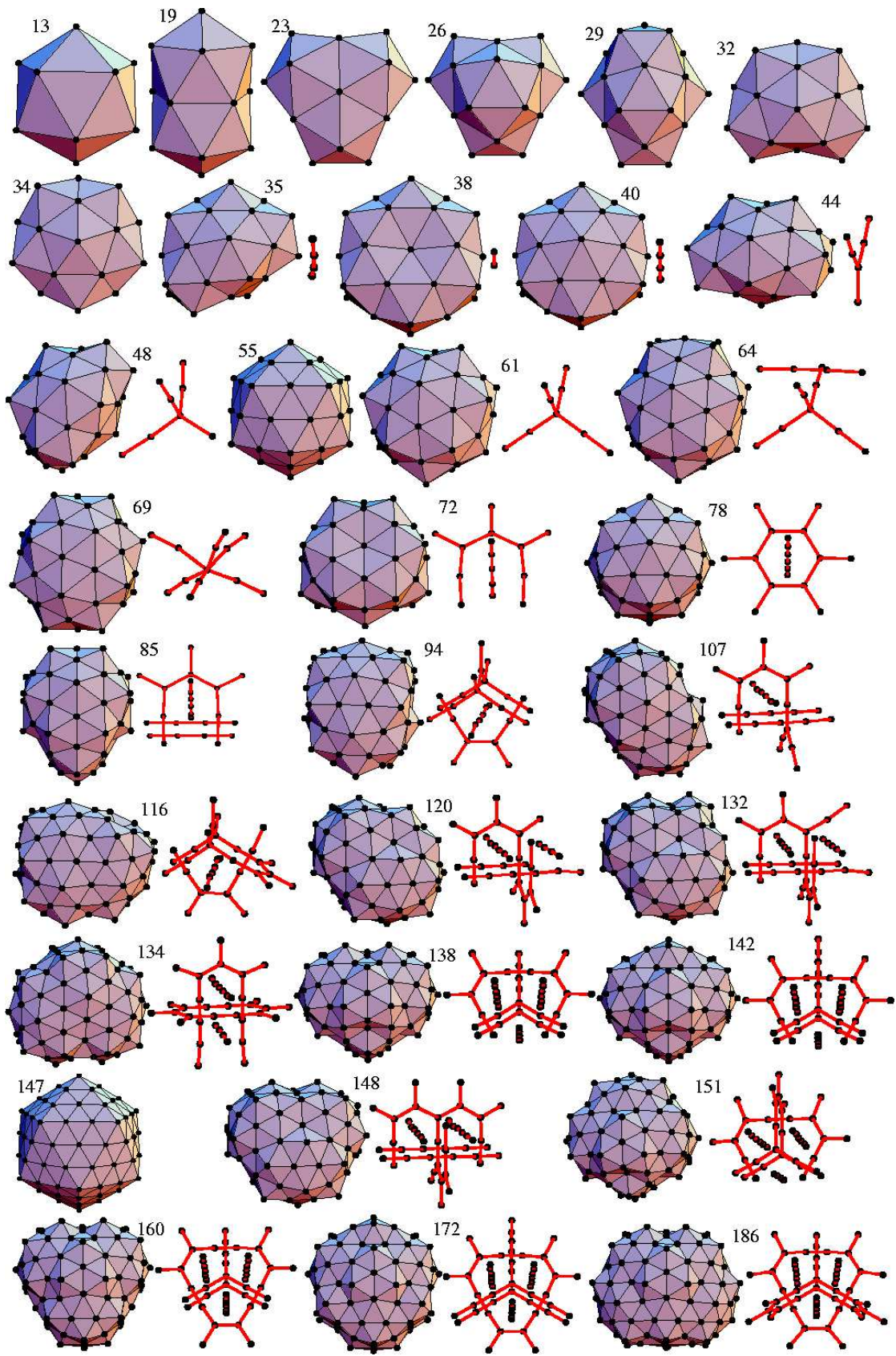}
\caption{\label{fig:structures} A selection of the putative global minima.
For the larger clusters the disclination network associated with a structure
is depicted to its right.}
\end{figure*}

The energies and point groups for the putative global minima are 
given in Table \ref{table:gmin}.
Point files are available online at the Cambridge Cluster Database.\cite{Web}
The energies of the global minima are represented in Figure \ref{fig:EvN} 
in such a way that makes particularly stable clusters stand out. 
A selection of clusters are depicted in Figure \ref{fig:structures}
that are either particularly stable or have some interesting structural
feature. Most of the intermediate sizes involve structures
similar to those in this figure, but normally with a somewhat incomplete 
surface structure.

\begin{table*}
\caption{\label{table:gmin}Energies and point groups (PG) of the putative 
global minima. The unit of energy is eV.}
\begin{ruledtabular}
\begin{tabular}{ccccccccccccccccccc}
$N$ & PG & Energy & & $N$ & PG & Energy & & $N$ & PG & Energy & & $N$ & PG & Energy & & $N$ & PG & Energy \\
\hline
   3 &  $D_{3h}$ &    -4.099029 & &   41 &  $C_{2v}$ &  -112.178113 & &   79 &  $C_s$    &  -227.336269 & &  117 &  $C_1$    &  -343.817038 & &  155 &  $C_s$    &  -461.506096 \\ 
   4 &  $T_d$    &    -6.242292 & &   42 &  $C_1$    &  -115.154425 & &   80 &  $C_s$    &  -230.318346 & &  118 &  $C_1$    &  -346.764630 & &  156 &  $C_{3v}$ &  -464.682681 \\ 
   5 &  $D_{3h}$ &    -8.607257 & &   43 &  $C_s$    &  -118.128134 & &   81 &  $C_s$    &  -233.343636 & &  119 &  $C_1$    &  -349.862568 & &  157 &  $C_1$    &  -467.796799 \\ 
   6 &  $O_h$    &   -11.257920 & &   44 &  $C_{2v}$ &  -121.260703 & &   82 &  $C_2$    &  -236.308999 & &  120 &  $C_2$    &  -353.034842 & &  158 &  $C_2$    &  -470.980712 \\ 
   7 &  $D_{5h}$ &   -13.860178 & &   45 &  $C_1$    &  -124.074147 & &   83 &  $C_1$    &  -239.278725 & &  121 &  $C_1$    &  -356.055763 & &  159 &  $C_s$    &  -474.154929 \\ 
   8 &  $C_{2v}$ &   -16.353276 & &   46 &  $C_s$    &  -127.238882 & &   84 &  $C_s$    &  -242.456890 & &  122 &  $C_2$    &  -359.102294 & &  160 &  $D_{3h}$ &  -477.328706 \\ 
   9 &  $C_{2v}$ &   -19.063671 & &   47 &  $C_s$    &  -130.228220 & &   85 &  $C_{2v}$ &  -245.646464 & &  123 &  $C_1$    &  -362.088084 & &  161 &  $C_1$    &  -480.408455 \\ 
  10 &  $C_{3v}$ &   -21.862051 & &   48 &  $C_{3v}$ &  -133.418980 & &   86 &  $C_s$    &  -248.606413 & &  124 &  $C_1$    &  -365.266998 & &  162 &  $C_s$    &  -483.573938 \\ 
  11 &  $C_{2v}$ &   -24.616176 & &   49 &  $C_s$    &  -136.111570 & &   87 &  $C_1$    &  -251.592963 & &  125 &  $C_1$    &  -368.426994 & &  163 &  $C_s$    &  -486.653798 \\ 
  12 &  $C_{5v}$ &   -27.780117 & &   50 &  $C_s$    &  -139.090832 & &   88 &  $C_s$    &  -254.736772 & &  126 &  $C_1$    &  -371.572700 & &  164 &  $C_2$    &  -489.818114 \\ 
  13 &  $I_h$    &   -31.278787 & &   51 &  $C_1$    &  -142.098016 & &   89 &  $C_s$    &  -257.685396 & &  127 &  $C_1$    &  -374.661505 & &  165 &  $C_s$    &  -492.896801 \\ 
  14 &  $C_{3v}$ &   -33.585594 & &   52 &  $C_1$    &  -145.091286 & &   90 &  $C_s$    &  -260.675290 & &  128 &  $C_1$    &  -377.808061 & &  166 &  $C_s$    &  -496.054542 \\ 
  15 &  $C_{2v}$ &   -36.321872 & &   53 &  $C_s$    &  -148.261944 & &   91 &  $D_3$  &  -263.877164 & &  129 &  $C_3$    &  -380.862378 & &  167 &  $C_1$    &  -499.123863 \\ 
  16 &  $C_s$    &   -39.039888 & &   54 &  $C_{5v}$ &  -151.376943 & &   92 &  $C_2$    &  -266.936839 & &  130 &  $C_1$    &  -383.890586 & &  168 &  $C_2$    &  -502.260998 \\ 
  17 &  $C_2$    &   -41.750455 & &   55 &  $I_h$    &  -154.612749 & &   93 &  $C_1$    &  -270.089293 & &  131 &  $C_1$    &  -387.030763 & &  169 &  $C_1$    &  -505.328765 \\ 
  18 &  $C_s$    &   -44.777004 & &   56 &  $C_{3v}$ &  -157.245282 & &   94 &  $C_{2v}$ &  -273.244125 & &  132 &  $C_2$    &  -390.138328 & &  170 &  $C_s$    &  -508.459278 \\ 
  19 &  $D_{5h}$ &   -48.182587 & &   57 &  $S_4$  &  -160.381251 & &   95 &  $C_s$    &  -276.204093 & &  133 &  $C_s$    &  -393.177026 & &  171 &  $C_s$    &  -511.527235 \\ 
  20 &  $C_{2v}$ &   -50.823659 & &   58 &  $C_s$    &  -163.479187 & &   96 &  $C_2$    &  -279.154395 & &  134 &  $C_2$    &  -396.274720 & &  172 &  $D_{3h}$ &  -514.656897 \\ 
  21 &  $C_1$    &   -53.470949 & &   59 &  $C_{2v}$ &  -166.601625 & &   97 &  $C_1$    &  -282.115807 & &  135 &  $C_s$    &  -399.361340 & &  173 &  $C_2$    &  -517.723146 \\ 
  22 &  $C_s$    &   -56.404346 & &   60 &  $C_{3v}$ &  -169.702701 & &   98 &  $C_1$    &  -285.257294 & &  136 &  $C_s$    &  -402.433300 & &  174 &  $C_1$    &  -520.801927 \\ 
  23 &  $D_{3h}$ &   -59.732308 & &   61 &  $T_d$    &  -172.787060 & &   99 &  $C_1$    &  -288.206364 & &  137 &  $C_s$    &  -405.526322 & &  175 &  $C_1$    &  -523.966150 \\ 
  24 &  $C_{2v}$ &   -62.350544 & &   62 &  $C_s$    &  -175.512306 & &  100 &  $C_1$    &  -291.309715 & &  138 &  $C_{2v}$ &  -408.694151 & &  176 &  $C_1$    &  -527.044129 \\ 
  25 &  $C_s$    &   -65.195739 & &   63 &  $C_1$    &  -178.554122 & &  101 &  $C_1$    &  -294.334721 & &  139 &  $C_s$    &  -411.773939 & &  177 &  $C_2$    &  -530.208354 \\ 
  26 &  $T_d$    &   -68.459809 & &   64 &  $C_{2v}$ &  -181.598713 & &  102 &  $C_1$    &  -297.418136 & &  140 &  $C_s$    &  -414.942360 & &  178 &  $C_1$    &  -533.271452 \\ 
  27 &  $C_{2v}$ &   -71.162534 & &   65 &  $C_1$    &  -184.466611 & &  103 &  $C_1$    &  -300.517946 & &  141 &  $C_s$    &  -418.022390 & &  179 &  $C_1$    &  -536.411855 \\ 
  28 &  $C_s$    &   -73.946989 & &   66 &  $C_1$    &  -187.498248 & &  104 &  $C_1$    &  -303.576439 & &  142 &  $C_{2v}$ &  -421.159084 & &  180 &  $C_1$    &  -539.475366 \\ 
  29 &  $D_{3h}$ &   -77.115629 & &   67 &  $C_1$    &  -190.610516 & &  105 &  $C_s$    &  -306.773455 & &  143 &  $C_s$    &  -424.234289 & &  181 &  $C_s$    &  -542.615132 \\ 
  30 &  $C_{2v}$ &   -79.792682 & &   68 &  $C_1$    &  -193.508419 & &  106 &  $C_1$    &  -309.860784 & &  144 &  $C_{2v}$ &  -427.293974 & &  182 &  $C_1$    &  -545.590132 \\ 
  31 &  $C_s$    &   -82.634873 & &   69 &  $C_3$    &  -196.696161 & &  107 &  $C_s$    &  -312.991830 & &  145 &  $C_1$    &  -430.398146 & &  183 &  $C_1$    &  -548.728221 \\ 
  32 &  $C_{2v}$ &   -85.678623 & &   70 &  $C_2$    &  -199.701093 & &  108 &  $C_s$    &  -315.964908 & &  146 &  $C_1$    &  -433.528391 & &  184 &  $C_2$    &  -551.764208 \\ 
  33 &  $C_{5v}$ &   -88.453125 & &   71 &  $C_s$    &  -202.700030 & &  109 &  $C_1$    &  -319.100302 & &  147 &  $I_h$    &  -436.702421 & &  185 &  $C_1$    &  -554.927247 \\ 
  34 &  $D_{5h}$ &   -91.447319 & &   72 &  $C_s$    &  -205.819266 & &  110 &  $C_1$    &  -322.261377 & &  148 &  $C_s$    &  -439.807282 & &  186 &  $C_2$    &  -558.089863 \\ 
  35 &  $C_{2v}$ &   -94.464663 & &   73 &  $C_s$    &  -208.794190 & &  111 &  $C_1$    &  -325.367188 & &  149 &  $C_1$    &  -442.741345 & &  187 &  $C_1$    &  -561.167833 \\ 
  36 &  $C_s$    &   -97.324952 & &   74 &  $C_s$    &  -211.818219 & &  112 &  $C_1$    &  -328.390553 & &  150 &  $C_s$    &  -445.909288 & &  188 &  $C_1$    &  -564.330048 \\ 
  37 &  $C_s$    &  -100.214276 & &   75 &  $C_s$    &  -215.088423 & &  113 &  $C_1$    &  -331.327112 & &  151 &  $C_{2v}$ &  -449.086382 & &  189 &  $C_1$    &  -567.390087 \\ 
  38 &  $D_{6h}$ &  -103.264168 & &   76 &  $D_{3h}$ &  -218.367742 & &  114 &  $C_s$    &  -334.422411 & &  152 &  $C_s$    &  -452.190934 & &  190 &  $C_2$    &  -570.537640 \\ 
  39 &  $C_{6v}$ &  -106.437242 & &   77 &  $C_{3v}$ &  -221.514884 & &  115 &  $C_1$    &  -337.598496 & &  153 &  $C_{2v}$ &  -455.278091 \\ 
  40 &  $D_{6h}$ &  -109.401628 & &   78 &  $D_{3h}$ &  -224.595480 & &  116 &  $C_s$    &  -340.759024 & &  154 &  $C_s$    &  -458.319723 \\ 
\end{tabular}
\end{ruledtabular}
\end{table*}

The only other study that has examined the structures of aluminium clusters
interacting with the present potential studied clusters with 
13, 43, 55 and 147 atoms.\cite{Sun98} For $N$=13, 55 and 147, Sun and Gong 
considered high symmetry structures and found the Mackay icosahedra 
to be lowest in energy, in agreement with the results here. For $N$=43, they
performed simulated annealing, generating a lowest-energy structure that is
$0.0562\,$eV above the global minimum, and which is in fact 
the second lowest-energy isomer. They also claimed that this cluster
had a `glassy' structure, but in common with the global minimum
it in fact has an ordered polytetrahedral structure.

The structures with from 3 to 13 atoms are those typically seen for 
isotropic potentials. Most are on a polytetrahedral growth sequence
leading to the 13-atom icosahedron, except for the six-atom octahedron 
and the 8-atom dodecehedron. Beyond $N=13$ growth occurs around the 
icosahedron with atoms being added above the faces and vertices of this structure 
in the so-called anti-Mackay overlayer.\cite{Doye95c} 
This growth maintains the  
polytetrahedral character of the clusters and leads to a series of 
interpenetrating 13-atom icosahedra at $N=19$, 23, 26, 29, 32 and 34. 
For example, the 34-atom structure involves seven interpenetrating icosahedra
with their centres in the shape of a pentagonal bipyramid.
This growth sequence is very similar to that for Lennard-Jones 
clusters\cite{Northby87}
although in a few instances slightly different sites in the 
anti-Mackay overlayer are occupied.

Beyond this size is where things become different from any previous system.
For example, for Lennard-Jones clusters there is a crossover at $N=31$ to the Mackay 
overlayer,\cite{Northby87} which leads to the 55-atom Mackay icosahedron. For model metal potentials,
even those that favour Mackay icosahedra, one often sees decahedral and close-packed structures 
at sizes away from the complete icosahedra.\cite{Doye98c} Instead, for these aluminium clusters
the polytetrahedral character of the global minima continues. However, it is not by continuation
of the anti-Mackay overlayer as for long-ranged Morse clusters,\cite{Doye97d} but
through structures involving ordered arrays of disclinations. Initially some of these structures
are similar to that for a modified form of the Dzugutov potential that was designed to produce
compact polytetrahedral clusters,\cite{Doye01d} but they then quickly diverge.

The 40-atom global minimum is a sixfold-symmetric equivalent of Al$_{34}$. It involves
a ring of six interpenetrating icosahedra with two interpenetrating Z14 
polyhedra along the symmetry axis. 
These latter two polyhedra give rise to the single disclination line
that passes through the structure. 
Similar structures have previously been found for Dzugutov 
clusters,\cite{Doye01a,Doye01d} and it is a small fragment of the 
Frank-Kasper Z phase.

Similar clusters are seen for $N=44$ and 48, 
but with a Z15 and Z16 polyhedron, respectively, 
lying slightly off-centre; these act as three-fold and 
four-fold nodes in the disclination networks.
Only for Al$_{48}$ does further growth occur on this cluster 
leading to the particularly stable 61-atom 
global minimum with tetrahedral symmetry.
This cluster was first identified for a long-ranged 
Morse cluster,\cite{Doye97d} and has since been found for clusters with a 
generalized set of metallic interactions,\cite{Cune00} and in experiments
on strontium clusters.\cite{Wang01}

Two exceptions to this dominance of polytetrahedral structures occur
at $N=54$ and 55, where a complete Mackay icosahedron and one with a vertex
missing atom are most stable. However, away from this magic number structures
based on Mackay icosahedra quickly become significantly higher in energy than 
the polytetrahedral structures.

The 64-atom global minimum is the first to involve not just a single network
of connected disclinations. It is based on Al$_{61}$ but it also
has a single linear disclination passing through the clusters. Such
disclination lines are a common motif for the larger clusters.
At $N=69$ the global minimum has an unusual structure with a 19-coordinate central
atom that acts as a node for six disclinations going out from it.

\begin{figure}
\includegraphics[width=8.4cm]{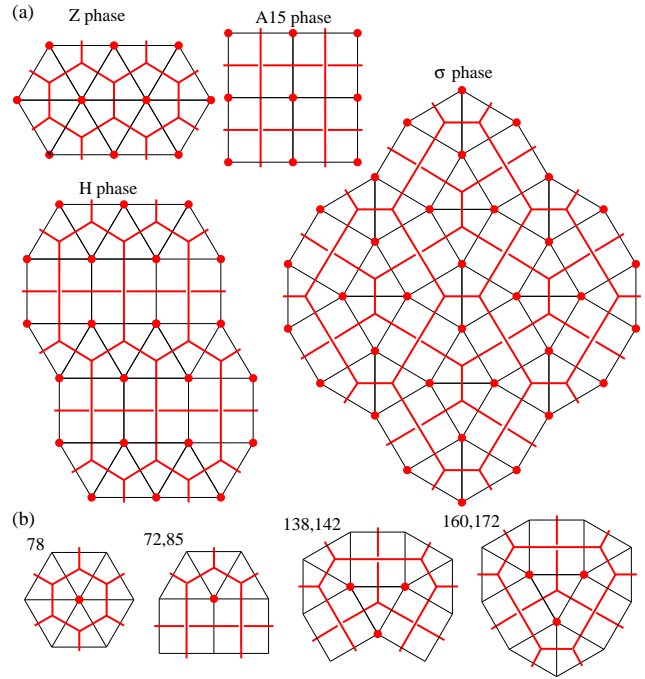}
\caption{\label{fig:ST} (a) The four square-triangle Frank-Kasper phases. 
The diagrams show a portion of the square-triangle tiling (thin black lines) 
for these phases. 
(b) Some of the square-triangle motifs that occur in the clusters in 
Fig.\ \ref{fig:structures} labelled by their sizes. 
The disclination networks (thick red lines and dots) associated with 
these tilings are also depicted.
The red dots represent disclinations coming out of the page. The disclination lines
that cross in the square tiles are at different heights in the crystal.}
\end{figure}

Starting at $N$=72 a new class of polytetrahedral clusters becomes most stable.
They are all based in some way on a set of Frank-Kasper phases, which are layer
structures\cite{Shoemaker} and can be considered as a 
tiling of squares and triangles that are suitably decorated with atoms.
The four simplest crystalline tilings are represented in 
Figure \ref{fig:ST}(a). 
The four Frank-Kasper phases
correspond to the four possible ways in which a point can be surrounded by 
squares and triangles. The Z phase has six triangles around each vertex, the 
A15 phase four squares and the H and $\sigma$ phases correspond to the two ways
that two squares and three triangles can share a common vertex.

Overlaid on the tilings in Figure \ref{fig:ST}(a) are the associated 
disclination networks. All have planar arrays of disclinations that are 
mediated by Z14 and Z15 polyhedra 
and that are threaded by linear disclinations that 
are perpendicular to the plane of the tiling. 
A number of the global minima can be
understood simply in terms of finite square-triangle tilings, as illustrated
in Figure \ref{fig:ST}(b). Al$_{78}$ has a disclination network 
that is part of the Z phase; Al$_{72}$ and Al$_{85}$ correspond to the 
H phase, and $N=138$, 142, 160 and 172 to a $\sigma$-like tiling.
The latter slightly deviates from the $\sigma$ phase, because, although 
the local arrangement of squares and triangles around all the
vertices in the tiling is the same as for the $\sigma$ phase, 
the central triangle only shares edges with squares. 
In fact this motif is common in models of dedecagonal 
quasicrystals.\cite{Stampfli88,Dzugutov93,Roth98}

The other larger global minima can be basically thought of in terms 
of these tilings, but with some deviations in the disclination networks
at the cluster surface. So $N=107$, 120, 132, 134 and 148 have H-like 
disclination networks, where the Al$_{85}$ tiling has been extended 
downward for $N=134$ and across for $N=120$, 132 and 148.
Similarly, $N=94$, 116, 151 and 186 also have $\sigma$-like tilings. 
All the $\sigma$-like structures are heading towards or based upon Al$_{160}$.

The deviations from the square-triangle structures involve breaking the planarity
of the disclination networks to connect disclinations
in different layers of the structure. The most common is to link two parallel disclinations
that would otherwise have simply exited the cluster.
This connection gives a disclination pattern similar to that for a three triangle tiling,
but in a plane perpendicular to the usual plane of the disclination network.
For example, the only difference between the structures at $N=94$ and 116, and 138 and 151
is the addition of one of these linkages.  Similarly, by the addition of two of these 
linkages one can generate Al$_{186}$ from Al$_{160}$. This motif is also
present in the structures at $N=120$, 132, 134 and 148.
The other linkage is similar and involves linking what would otherwise have been 
three-fold nodes in the disclination network, thus generating Z16 nodes. 
This type of connection occurs for $N$=94 and 116.

It is noticeable that the variations in the energy with respect to the average for clusters
with more than 100 atoms are relatively small (Fig.\ \ref{fig:EvN}). 
This is because there are lots of ways of creating
polytetrahedral packings. This behaviours is unlike what 
occurs, say, for structures based 
on the Mackay icosahedra, where, between the magic numbers, 
structures with incomplete outer layers
give rise to a substantial energy variation. Even the most stable large
polytetrahedral packings, the $\sigma$-like structures centred around $N=166$, 
only give rise to a broad minimum in Fig.\ \ref{fig:EvN}. 
The structure at the minimum, Al$_{166}$, is in fact intermediate between 
the 160- and 172-atom structures depicted in 
Fig.\ \ref{fig:structures} with three of the six disclinations extended.

\section{\label{sect:Al61}Al$_{61}$: a case study}

\begin{figure}
\includegraphics[width=8.4cm]{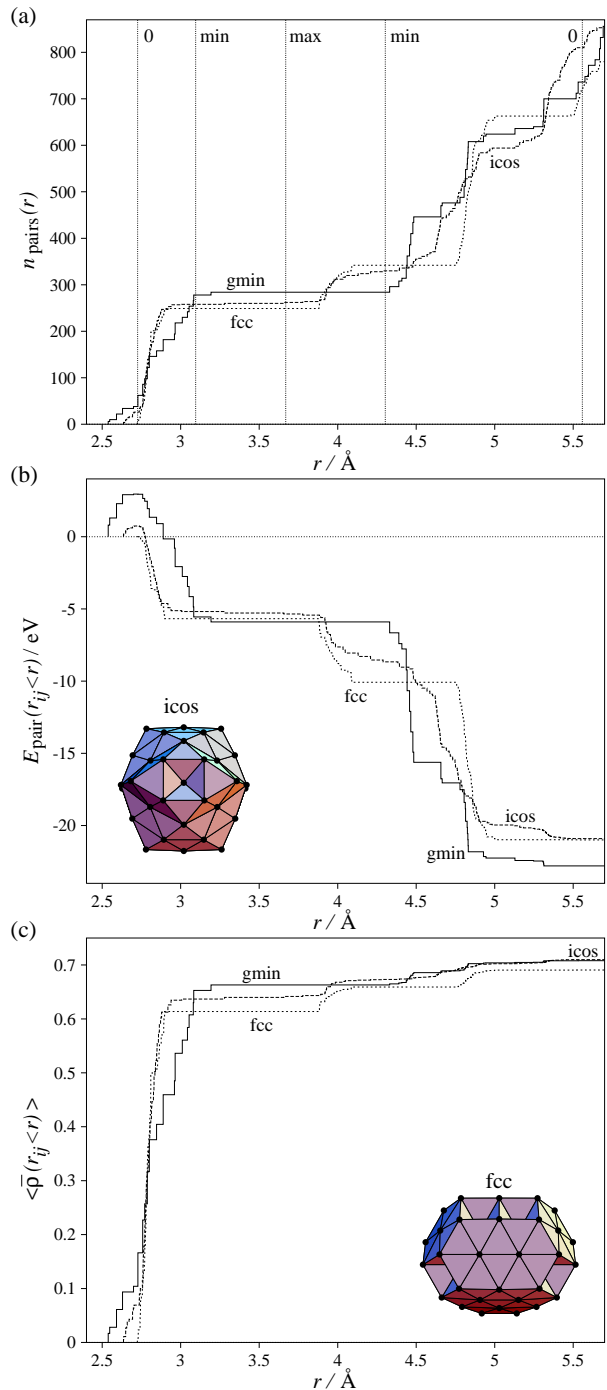}
\caption{\label{fig:Edecomp}A comparison of the properties
of the global minimum (gmin), the lowest-energy icosahedral structure (icos)
and the lowest-energy fcc structure for $N=61$. The latter two structures
are illustrated in the corners of (b) and (c), respectively.
(a) $n_{\rm pairs}(r)$. (b) $E_{\rm pair}(r_{ij}<r)$. 
(c) $\langle \bar\rho(r_{ij}<r)\rangle$. 
In (a) vertical lines have been added corresponding to the values of $r$ where 
the pair potential passes through zero, is a minimum or a maximum, and
goes to zero at the cutoff, as labelled.}
\end{figure}

To try to unravel why the aluminium
glue potential favours polytetrahedral structures, 
in this section I take Al$_{61}$ as a case study 
and compare the energetics of some competing structural forms
(Table \ref{table:Edecomp}). 
The structures I consider are the polytetrahedral global
minimum (Fig. \ref{fig:structures}), an icosahedral structure
that can be formed by adding a six-atom overlayer to the 55-atom 
Mackay icosahedron and an fcc cluster. The latter two are illustrated
in the insets to Fig.\ \ref{fig:Edecomp}.

The decomposition of the potential energy into glue and pair 
contributions shows 
(Table \ref{table:Edecomp}) that the global minimum is stabilized
by a significantly lower $E_{\rm pair}$. This is partially offset by 
a higher $E_{\rm glue}$ than the icosahedral structure.
Figures \ref{fig:Edecomp}(a) and (b) enable us to understand why 
the pair energy 
is lowest for the polytetrahedral clusters, where (a) depicts 
$n_{\rm pairs}(r)$, the number of pairs of atoms that are
separated by less than $r$, and (b) 
$E_{\rm pair}(r_{ij}<r)$, the contribution to $E_{\rm pair}$ from
pairs of atoms that are separated by less than $r$.

The contribution to $E_{\rm pair}$ from the 
nearest neighbour shell (defined as those pairs with a separation less
than 3.5\AA) can be decomposed into two terms
\begin{equation}
E_{\rm pair}^{\rm nn}= n_{\rm nn} \phi(\langle r_{\rm nn}\rangle) + 
\Delta E_{\rm pair}^{\rm dist}, 
\end{equation}
where 
$n_{\rm nn}$ is the number of nearest neighbours,
$\langle r_{\rm nn}\rangle$ is the
average separation of pairs in the nearest-neighbour shell.
The first term is the $E_{\rm pair}^{\rm nn}$ value that 
would result if all the $n_{\rm nn}$ nearest neighbours had the same 
pair separation, and the second term, $\Delta E_{\rm pair}^{\rm dist}$, is the 
energy cost associated with the nearest-neighbour pair distances having 
a distribution of values.\cite{strain}

As is well-known icosahedral clusters have larger values of $n_{\rm nn}$ than 
fcc clusters, because Mackay icosahedra are quasispherical and only
have high-coordinate $\{111\}$ faces, whereas the most spherical fcc 
clusters must have a significant proportion of lower-coordinated $\{100\}$ 
faces. This more favourable surface energy is the reason why icosahedral 
clusters are often lowest in energy for small clusters interacting with 
pair potentials. The polytetrahedral global minimum has an even higher value 
of $n_{\rm nn}$ than the icosahedral structure because of the higher 
coordination number for the interior atoms that lie on disclinations, 
and the densely packed surface. 

As anticipated in Section \ref{sect:potential}, $\langle r_{\rm nn}\rangle$
is significantly shorter than the distance corresponding to 
the first minimum in the pair potential---some nearest-neighbour pairs in
the polytetrahedral and icosahedral structures even have positive 
pair energies. $\langle r_{\rm nn}\rangle$ also has a significant structural
dependence. For the icosahedral and fcc structures, $\langle r_{\rm nn}\rangle$
is contracted with respect to $r_{\rm nn}^{\rm xtal}$, as is usual for 
metal clusters,\cite{Soler00} in order to try to increase the 
$\bar\rho$ values of the surface atoms.
By contrast, $\langle r_{\rm nn}\rangle$ for the global minimum
is actually slightly larger than $r_{\rm nn}^{\rm xtal}$. 
This is not that there is no contraction at the surface, but rather that this 
contraction is with respect to the appropriate bulk Frank-Kasper phase.

\begin{table}
\caption{\label{table:Edecomp}
The properties of 
a series of 61-atom structures, namely the global minimum (gmin), 
the lowest-energy icosahedral (icos) and lowest-energy fcc structure.
$\sigma_{\bar\rho}$ is the standard deviation in the values of $\bar\rho_i$,
and $\sigma_r$ is the standard deviation in the nearest-neighbour distances. 
$n_{\rm bulk}$ is the number of atoms in the interior of the cluster, 
where an atom is deemed to be on the surface if its coordination number 
is ten or fewer. 
$E_{\rm bulk}$ and $E_{\rm surf}$ are the average energies of the atoms
in the interior and on the surface of the cluster, respectively. 
All other quantities are defined in the text.
}
\begin{ruledtabular}
\begin{tabular}{cccc}
                                & gmin       & icos        & fcc  \\ 
\hline
point group                     & $T_d$    & $C_{2v}$ & $C_{3v}$ \\
$E$ / eV                     & -172.787 & -171.697 & -170.975 \\
$E_{\rm pair}$ / eV             &  -22.795 &  -20.901 &  -20.991 \\
$E_{\rm glue}$ / eV             & -149.992 & -150.796 & -149.984 \\
$E_{\rm pair}^{\rm nn}$ / eV    &   -5.899 &   -5.284 &   -5.680 \\
$n_{\rm nn}$                    & 284      & 260      & 249      \\
$\langle r_{\rm nn}\rangle$     &   2.8536 &  2.7973  &   2.7974 \\
$\sigma_r$                      &   0.1565 &  0.0742  &   0.0442 \\
$\Delta E_{\rm pair}^{\rm dist}$ / eV & 4.817 & 0.997 &    0.342 \\
$\langle \bar\rho\rangle$       &   0.7078 &  0.7098  &   0.6906 \\
$\sigma_{\bar\rho}$             &   0.2145 &  0.2035  &   0.1904 \\
$\Delta E_{\rm glue}^{\rm dist}$ / eV & 6.052 & 5.338 &   5.159  \\
$n_{\rm bulk}$                  &  17      &  14      &  13      \\
$E_{\rm bulk}$ / eV atom$^{-1}$ &   -3.104 & -3.119   &  -3.186  \\
$E_{\rm surf}$ / eV atom$^{-1}$ &   -2.728 & -2.724   &  -2.699  \\
\end{tabular}
\end{ruledtabular}
\end{table}

Given the large value of $n_{\rm nn}$ for the global minimum and that
$\langle r_{\rm nn}\rangle$ lies closest to the potential minimum, it is not 
so surprising that the global minimum has the lowest value of 
$E_{\rm pair}^{\rm nn}$.
However, it is only marginally lower than that for the fcc cluster, which
in turn is lower than the icosahedral cluster (Figure \ref{fig:Edecomp}(b)) 
even though it has fewer nearest neighbours.
These effects can be understood by taking account of the 
distribution of $r_{ij}$ values. 
As $\phi(r)$ has positive curvature in this region, 
$\Delta E_{\rm pair}^{\rm dist}$ must be positive. 

As is evident from Figure \ref{fig:Edecomp}(a) and the $\sigma_r$ values
in Table \ref{table:Edecomp}, the global minimum has the widest distribution 
of nearest-neighbour distances, and hence by far the largest 
$\Delta E_{\rm pair}^{\rm dist}$. By contrast the fcc cluster has the 
narrowest distribution and smallest $\Delta E_{\rm pair}^{\rm dist}$. 
This explains why $E_{\rm pair}^{\rm nn}$ for the fcc cluster is actually lower
than that for the icosahedral cluster and only just above that for the 
global minimum. 
The large value of $\sigma_r$ for the global minimum reflects 
the distortions of the tetrahedra away from regularity that are necessary 
in any polytetrahedral packing. 

$E_{\rm pair}^{\rm nn}$, however, only represents roughly a quarter of
the total pair energy;
the dominant contribution energy comes from next neighbours.
There are a number of possible configurations for such neighbours. 
The nearest next neighbours correspond to those at opposite vertices of 
an octahedron, which if regular corresponds
to a separation of $\sqrt{2}\,r_{\rm nn}=1.4142\,r_{\rm nn}$. These
configurations occur most frequently in the fcc clusters, also 
occur in the twenty fcc tetrahedra that make up the Mackay icosahedron, 
but do not occur at all in a polytetrahedral structure. 

The next set of pair distances correspond to atoms at opposite ends
of a trigonal bipyramid,
which if it is made of two regular tetrahedra occurs at 
$2\sqrt{2/3}\,r_{\rm nn}=1.6330\,r_{\rm nn}$. These configurations are
absent from fcc clusters, occur around the fivefold axes of the Mackay icosahedra
and also across the twin planes between the strained fcc tetrahedra that
make up the icosahedron and are most prevalent in polytetrahedral 
clusters.

Then, there are pairs at the extreme ends of a planar rhombus, which if 
it consists of two equilateral triangles would be at a separation of 
$\sqrt{3}\,r_{\rm nn}=1.7321\,r_{\rm nn}$. However, these pairs, which are 
most common in the fcc structures due to the close-packed planes, 
are less important to the energetics because, of the three sets of 
configurations, they are the furthest from 
the second minimum in the pair potential.

Given that $r_{\rm min2}=1.509\, r_{\rm nn}^{\rm xtal}$, one might
expect from the above ideal configurations that the octahedral pairs
would lie slightly closer to the second potential minimum than those
pairs at opposite ends of a trigonal bipyramid. However, the contraction
of $\langle r_{\rm nn} \rangle$ with respect to the crystal takes 
the octahedral distances further away from this minimum. 
Furthermore, the wide distribution of nearest neighbour distances in 
the polytetrahedral structures also leads to a wide variation in the
dimensions of the trigonal bipyramids.
As can be seen from Figure \ref{fig:Edecomp}(a) a significant fraction of 
these polytetrahedral pairs for the global minimum
lie closer to the pair potential minimum than the octahedral distances.
These effects are responsible for the substantially lower pair energy of the 
global minimum (Figure \ref{fig:Edecomp}(b)). 
The trigonal bipyramids present in the icosahedral structure also
cause the difference in the total pair energy between it
and the fcc structure to narrow, compared to that for 
$E_{\rm pair}^{\rm nn}$. 

Similar to the analysis for $E_{\rm pair}^{\rm nn}$ the glue energy can
be written as
\begin{equation}
E_{\rm glue}=N U(\langle \bar\rho \rangle) +\Delta E_{\rm glue}^{\rm dist}
\end{equation}
where the first term is the glue energy that would result if all atoms
had the same value of $\bar\rho$ 
($\langle \bar\rho \rangle$ is the average value of $\bar\rho$), 
and the second term is the 
correction due to the $\{\bar\rho_i\}$ having a distribution of values.

We can examine the role of different pair distances on the glue energy 
by following $\langle \bar\rho(r_{ij}<r) \rangle$, the contribution to 
$\langle \bar\rho \rangle$ from pairs of atoms that are separated by 
less than $r$. It is clear from Figure \ref{fig:Edecomp}(c) that, 
as anticipated, the major contribution to $\langle \bar\rho \rangle$
comes from nearest neighbours, and the values of 
$\langle \bar\rho(r_{ij}<r) \rangle$ after the nearest-neighbour shell
reflect the relative values of $n_{\rm nn}$.
$\langle r_{\rm nn}\rangle$ is also important. 
The average contribution to $\langle \bar\rho\rangle$ from each 
nearest neighbour in the global minimum is less because of the 
longer $\langle r_{\rm nn}\rangle$. 
Indeed, if the three structures had the same $\langle r_{\rm nn}\rangle$ the
polytetrahedral structure would have comfortably the lowest glue energy. 
The differences in $\sigma_r$ have little effect, 
because the function $\rho(r)$ is approximately
linear for the relevant range of $r$ (Figure \ref{fig:potential}(a)).

As $\rho(r)$ is a monotonically decreasing function, the next 
largest contribution to $\langle \bar\rho \rangle$ comes 
from octahedral next neighbours. 
All other pairs have little effect on the relative values of 
$\langle \bar\rho \rangle$ for the different structures because
by these distances $\rho(r)$ is small in magnitude. 

The effect of next-nearest neighbours is to cause $\langle \bar\rho \rangle$
for the icosahedral structure to just overtake that for the global minimum,
and similarly to diminish the difference between 
$\langle \bar\rho \rangle$ for the global minimum and the fcc structure, 
compared to the relative values of $\langle \bar\rho(r_{ij}<r) \rangle$ 
after the nearest-neighbour shell.

The ordering of the $\langle \bar\rho \rangle$ values agrees with that for 
$E_{\rm glue}$.  However, although the global minimum has a similar value
of $\langle \bar\rho \rangle$ to that for the icosahedral structure, its glue 
energy is only just greater than that for the fcc structure. This effect
can be explained by taking into account the distribution of $\bar\rho_i$ values.
As $U(\bar\rho)$ has positive curvature, $\Delta E_{\rm glue}^{\rm dist}$
must be positive, and is in fact largest for the global minimum because it
has the greatest variance in $\bar\rho_i$ values (Table \ref{table:Edecomp}).
To a first approximation the distribution of $\bar\rho_i$ values is bimodal
with similar values for surface atoms, and similar values for interior atoms.
As the polytetrahedral cluster has the most atoms in the interior of the 
cluster due to the high coordination numbers, it has the largest 
$\sigma_{\bar\rho}$. 
 
As we have pointed out above the value of $\langle r_{\rm nn}\rangle$ 
is important for the values of both $E_{\rm pair}$ and $E_{\rm glue}$,
so it is important to understand what determines $\langle r_{\rm nn}\rangle$.
At $\langle r_{\rm nn}\rangle$ the decrease in the glue energy by uniformly 
shrinking the cluster is exactly balanced by the increase in the pair energy.
This energy balance occurs at larger $\langle r_{\rm nn}\rangle$ for the
global minimum for a number of reasons. Firstly, 
the decrease in the glue energy as the cluster is shrunk 
(this increases the $\bar\rho$ values of the surface atoms)
is smallest for the global minimum because it has the fewest number 
of atoms in the surface of the cluster and has some atoms with 
high coordination number in the cluster interior (see below).
Secondly, the non-linear increase in the pair energy as the cluster
is shrunk is largest for the global minimum. This mainly stems from
the nearest-neighbour contribution and reflects both the greater number
of nearest neighbours for the global minimum and its greater $\sigma_r$.

The average nearest-neighbour distance also shows a systematic dependence 
on the coordination number. For energetic (see Section \ref{sect:potential}) 
and geometric reasons the atoms with higher coordination numbers have larger 
average nearest-neighbour distances. For example, the nearest neighbours
of the sixteen-coordinate atom in the centre of the global minimum
are at an average separation of 2.933\,\AA, 
whereas the corresponding distance for the 12-coordinate atoms is 2.725\,\AA.

Although the analysis in this section has been done here for a 
single case study,
similar analyses for other cluster sizes have revealed the same 
principles at work.

\section{\label{sect:large}Larger clusters}

Although polytetrahedral clusters are dominant in the size range considered
in the last section, one would expect that clusters at sufficiently large size 
would adopt the bulk structure, which for aluminium is fcc. This
assumes that the current glue potential can correctly reproduce the bulk structure.
However, to the best of my knowledge, the relative energetics for different crystal structures
have not been compared for this potential; for a somewhat similar aluminium 
empirical potential the fcc crystal is only marginally lower in energy than 
an A15 Frank-Kasper phase.\cite{Mishin99}

\begin{figure}
\includegraphics[width=8.4cm]{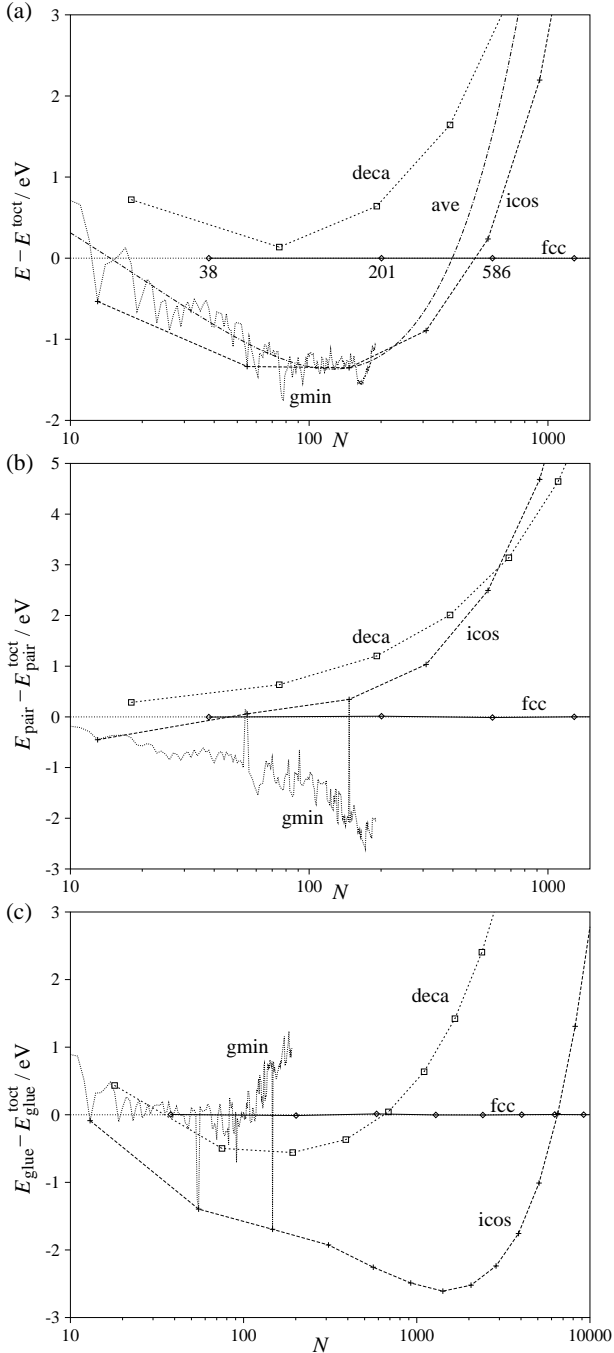}
\caption{\label{fig:E_large} A comparison of the relative values of 
(a) $E$, (b) $E_{\rm pair}$ and (c) $E_{\rm glue}$ for the 
global minima for $N\le190$ (gmin), and sequences of fcc truncated octahedra
with hexagonal $\{111\}$ facets (fcc), Mackay icosahedra (icos) and
Marks decahedra (deca). In each graph the energies are measured relative to fits
to those for the truncated octahedra, where 
$E_{\rm toct}=-3.360N+1.890N^{2/3}+1.170N^{1/3}-0.267$,
$E_{\rm toct}^{\rm glue}=-2.657N+0.456N^{2/3}+1.096N^{1/3}+0.621$ and
$E_{\rm toct}^{\rm pair}=-0.703N+1.435N^{2/3}+0.068N^{1/3}-0.873$.
In (a) $E_{\rm ave}$, the fit to energies of the global minima used in
Fig.\ \ref{fig:EvN}, has also been plotted.
}
\end{figure}

That for Al$_{61}$, and all the other clusters we examined in detail, $E_{\rm bulk}^{\rm fcc}$ is lower 
than that for the polytetrahedral and icosahedral clusters (Table \ref{table:Edecomp})
suggests that the system will converge to the correct bulk structure. 
To further understand the size evolution of the cluster structure I have
reoptimized a series of Mackay icosahedra, fcc truncated octahedra and Marks decahedra, which 
are usually the optimal shapes for these morphologies.\cite{Raoult89a} The energies of these
sequences are compared in Figure \ref{fig:E_large}(a) along with the global minima obtained in the Section
\ref{sect:gmin}. I have not attempted to construct a sequence of larger polytetrahedral 
clusters, because one cannot easily deduce from the global minima 
how the polytetrahedral growth sequence should continue. Instead, we also plot $E_{\rm ave}$, the fit 
to the energies of the global minima.

From Figure \ref{fig:E_large}(a) it is clear that decahedral structures are never competitive 
for this potential. This is slightly unusual as it is often found that for materials
which are close-packed in bulk, there is a progression from icosahedral to decahedral to 
close-packed structures as the size increases. Instead, the icosahedral and fcc lines cross
at $N\approx 520$ suggesting that beyound this size fcc clusters will be lower in energy 
for the majority of sizes. The $E_{\rm ave}$ curve also seems to suggest that these fcc clusters
will be global minima, but one has to be somewhat careful. Firstly, one should not necessarily trust 
this function outside the size range to which it has been fitted. Secondly, it represents the 
average energy of the polytetrahedral global minima, whereas the other 
curves in Figure \ref{fig:E_large}(a) are lower bounds to the energies of 
the clusters of that type. For example, $E_{\rm ave}$ only dips below the interpolated energies of 
the complete Mackay icosahedra for a small range of $N$, even though 
polytetrahedral structures are the global 
minima for the vast majority of the size range considered in Section \ref{sect:gmin}.

To get a better idea of the causes of this size evolution 
I have plotted in Figures \ref{fig:E_large}(b) and (c) the glue and 
pair contribution to the 
energies of these sequences of structures. It is immediately clear, 
in agreement with the analysis of Al$_{61}$, 
that the polytetrahedral global minima are stabilized by their particularly low 
pair energies. Furthermore, the global minima that are based on Mackay icosahedra at $N=54$--55 
and 147 clearly stand out from this trend.

It is also clear that the crossover from icosahedral to fcc structures is driven by 
the increasing advantage provided by its lower pair energy. By contrast, the 
glue energy of the icosahedron only becomes greater than that for the truncated octahedra
at $N\approx 6500$. $E_{\rm glue}$ more strongly reflects the greater number of nearest 
neighbours associated with the Mackay icosahedra and is less sensitive to the strain inherent to 
the icosahedra. By contrast, as noted in section \ref{sect:gmin} and as is usual for 
pair potentials,\cite{Doye97d} the nearest neighbour contribution
to the pair energy is reduced due to the strain in the pair distances for the icosahedra.

\section{\label{sect:comp}Experimental and theoretical comparisons}

Although I have focussed on the general principles that can be drawn 
from the current results, 
given that the potential has been successfully used in a wide 
variety of applications, with in some cases excellent agreement with 
experimental data,\cite{Sandberg02} one would hope that this system
would also be a reasonable model for real aluminium clusters.
However, one should remember that to correctly predict the energetics 
of cluster structure requires that the potential can successfully 
describe a wide variety of surface and bulk properties for a variety 
of structural types, and so is an extremely stringent test
of a potential.

Although aluminium clusters have been much studied experimentally,
there is little reliable information about the geometric structure 
in the size range of the global minima considered in Section \ref{sect:gmin}.
The difficulty is that features in the experimental data
can originate from the electronic shell structure
rather than the geometric structure.
Indeed, following the successful interpretation of many of the properties 
of alkali metal clusters in terms of the electronic shell structure,
aluminium, a nearly free-electron metal, was a natural system to 
see how widely applicable these ideas might be. 
Indications of this electronic shell structure were found
from the ionization potentials,\cite{Schriver90}
the magic numbers in mass spectra,\cite{Schriver90,Pellarin93}
cluster ion mobilities\cite{Jarrold93} and
photoelectron spectroscopy\cite{Li98,Akola99,Akola00}. 

However, a significant number of the features in these
properties could not be explained thus, and so these were 
usually attributed to the effects of the geometric structure.
Noteworthy of these is the magic number at $N=55$, whose
most obvious interpretation is in terms of a Mackay icosahedron.
No convincing explanations, however, have been given for the other 
`geometric' features. For example, except for $N=55$, the magic numbers do not 
correspond to those of any of the usual structural forms for clusters,
but neither do they correspond to the new polytetrahedral magic numbers
found here.

Theoretically, aluminium clusters in this size range 
have been recently studied using two empirical 
potentials, namely the Murrell-Mottram\cite{Lloyd98,Lloyd02} 
and Gupta\cite{Bailey03} forms.
For both these system icosahedral structures predominate for the 
global minima, although with fcc and decahedral structures competing 
at sizes intermediate between the complete Mackay icosahedra.
I have also reoptimized a series of icosahedral, decahedral, fcc and
polytetrahedral clusters for the potential of Mishin {\it et al.},
whose $\phi(r)$, $\rho(r)$ and $U(\bar\rho)$ functions resemble those for the 
current glue potential and were fit to a wide range of bulk data.\cite{Mishin99}
In this case polytetrahedral structures again predominate.

There have also been many electronic structure
calculations of small aluminium clusters using a variety of 
methods.\cite{Jones91,Yi91,Cheng91,Debiaggi92,Yang93,Akola98,Akola99,Akola00,Ahlrichs99}
For small clusters there has been some success in identifying the 
structure for $N<20$ by comparing calculated photoelectron spectra
with experiment.\cite{Akola99,Akola00}
However, for somewhat larger clusters 
there is little consensus among the results, partially because 
the computational expense prevents one rigorously searching for the global 
minimum---usually only a selection of clusters are compared---but also 
because different methodologies simply give different results.
For example, there is disagreement over whether for Al$_{55}$ the
Mackay icosahedron\cite{Yang93} or the fcc 
cuboctahedron\cite{Cheng91,Ahlrichs99,Akola00} is lower in energy,
and still others have found disordered structures to be lower in energy 
than both.\cite{Yi91}
A similar range of results have been found for 
Al$_{147}$.\cite{Debiaggi92,Ahlrichs99}

For large aluminium clusters there are clear experimental indicators
of the geometric structure. For $N>250$ (up to at least $N\approx 10\,000$)
there are set of regular spaced magic numbers in the mass 
spectra.\cite{Lerme92,Martin92,Pellarin93,Baguenard}
Initially, this result was interpreted in terms of electronic-shell 
effects,\cite{Lerme92,Pellarin93} but Martin {\it et al.} soon 
showed that the magic numbers were in fact due to octahedra, with each
additional magic number corresponding to the addition of a complete overlayer
to a single face of the octahedra.\cite{Martin92,Naher93}
Growth simulations further added to the plausibility of this 
interpretation.\cite{Valkealahti94,Valkealahti95,Valkealahti98}

For large clusters one should always ask whether the observed structures
occur because they are most stable or because of the growth kinetics.
This is especially the case here, because the most stable fcc clusters are
usually more spherical truncated octahedra,\cite{van89} rather than octahedra.
Indeed as the life times of the clusters are increased, the magic
numbers disappear,\cite{Baguenard} indicating their kinetic origin.
However, it is probably reasonable to conclude that fcc structures are
most stable in this size range, even if the appearance of octahedra
is a kinetic effect that results from the nucleation of a new face 
being slower than the subsequent growth. 

My estimate that fcc structures become lowest in energy beyond $N\approx 500$ 
is therefore reasonably close to experiment. This contrasts with 
the Gupta potential where fcc structures only become more stable 
beyond $N\approx 1250$ and the Murrell-Mottram potential where the
crossover occurs at even larger sizes.\cite{Turner00} 
However, one should remember that these energetic crossover sizes
are only strictly valid at $T=0$K. If one compares the relative free 
energies, the crossover sizes can be sensitively dependent on the
temperature, mainly due to the differences in the vibrational entropy
of the competing structures.\cite{Doye01b,Doye02b}

\section{\label{sect:conc}Conclusion}
I have found that the global minimum of aluminium clusters described by
a glue potential to be predominantly polytetrahedral for $N<160$. The
structure of the larger clusters can be described in terms of their 
disclination networks, which resemble those of the square-triangle 
Frank-Kasper phases. This result is somewhat surprising given that  ordered 
polytetrahedral structures are only found for alloys in bulk, but it 
reflects the greater tendency for polytetrahedral order in clusters.
The bulk behaviour should of course emerge for sufficiently large clusters,
and our results suggest that for $N>500$ clusters 
with the bulk fcc structure predominate.

An important aspect of this work has been to relate the observed structures
back to the interatomic interactions. Polytetrahedral structures are 
favoured for a combination of reasons. Firstly, the clusters
present surfaces with a low surface energy and high average cooordination 
number, so will be potentially favoured at small sizes.
Secondly, the glue function, which gives rise to most of the 
cohesive energy of the clusters, is relatively insensitive to the 
strain in the pair distances and the high coordination numbers 
that are inevitable in polytetrahedral structures.
Thirdly, as with the modified Dzugutov potential,\cite{Doye01d} the 
two minima in the pair potential most closely match the distances in 
polytetrahedral structures.

As the first two of these reasons are true for most metal potentials, 
it is not so surprising that there is experimental evidence of polytetrahedral 
structures for monatomic cobalt,\cite{Dassenoy00} 
strontium,\cite{Wang01} and perhaps some rare-earth\cite{Brechignac93} 
clusters. Nor would it be surprising if the number of theoretical and 
experimental examples were to increase.
My results further\cite{Doye01d} illustrate the types of polytetrahedral 
structures that one might expect for such clusters, along with a series of 
magic numbers that will be useful in comparisons with experiment.

Whether the preference for polytetrahedral structures in the current model 
carries over to real aluminium clusters is not yet clear.
Because of the lack of easily interpretable experimental data
on the geometric structure of aluminium clusters with less than 190 atoms,
it is hard to draw any firm conclusions about the realism of the 
structures predicted by the glue potential in this size range. 
Moreover, {\it ab intio}
calculations add little further insight because of the lack of consensus.
However, comparisons to the experimental data for larger clusters suggest
that the glue potential is more realistic than the other empirical 
potentials for which calculations have been performed.

The properties of these clusters can also provide useful insights into 
the behaviour observed for this potential in other contexts. For example,
studies on aluminium nanowires have found polytetrahedral structures
for those wires with sufficiently small cross-section.\cite{Gulseren98} Some
of these involve only icosahedral local coordination, but others clearly
involve disclinations. For example, the wire that G\"{u}lseren {\it et al.}
denote as A7 has a single disclination running along the centre of the wire. 
Similarly, from the current results it should be unsurprising 
that there is strong polytetrahedral ordering in the aluminium melt
described by this potential.\cite{Liu01}


\begin{thebibliography}{89}
\expandafter\ifx\csname natexlab\endcsname\relax\def\natexlab#1{#1}\fi
\expandafter\ifx\csname bibnamefont\endcsname\relax
  \def\bibnamefont#1{#1}\fi
\expandafter\ifx\csname bibfnamefont\endcsname\relax
  \def\bibfnamefont#1{#1}\fi
\expandafter\ifx\csname citenamefont\endcsname\relax
  \def\citenamefont#1{#1}\fi
\expandafter\ifx\csname url\endcsname\relax
  \def\url#1{\texttt{#1}}\fi
\expandafter\ifx\csname urlprefix\endcsname\relax\def\urlprefix{URL }\fi
\providecommand{\bibinfo}[2]{#2}
\providecommand{\eprint}[2][]{\url{#2}}

\bibitem[{\citenamefont{Alonso}(2000)}]{Alonso00}
\bibinfo{author}{\bibfnamefont{J.~A.} \bibnamefont{Alonso}},
  \bibinfo{journal}{Chem. Rev.} \textbf{\bibinfo{volume}{100}},
  \bibinfo{pages}{637} (\bibinfo{year}{2000}).

\bibitem[{\citenamefont{Johnston}(2002)}]{Johnston02}
\bibinfo{author}{\bibfnamefont{R.~L.} \bibnamefont{Johnston}},
  \emph{\bibinfo{title}{Atomic and Molecular Clusters}}
  (\bibinfo{publisher}{Taylor \& Francis}, \bibinfo{address}{London},
  \bibinfo{year}{2002}).

\bibitem[{\citenamefont{Wales and Scheraga}(1999)}]{WalesS99}
\bibinfo{author}{\bibfnamefont{D.~J.} \bibnamefont{Wales}} \bibnamefont{and}
  \bibinfo{author}{\bibfnamefont{H.~A.} \bibnamefont{Scheraga}},
  \bibinfo{journal}{Science} \textbf{\bibinfo{volume}{285}},
  \bibinfo{pages}{1368} (\bibinfo{year}{1999}).

\bibitem[{\citenamefont{Wang et~al.}(2001)\citenamefont{Wang, Blaisten-Barojas,
  Roitberg, and Martin}}]{Wang01}
\bibinfo{author}{\bibfnamefont{G.~M.} \bibnamefont{Wang}},
  \bibinfo{author}{\bibfnamefont{E.}~\bibnamefont{Blaisten-Barojas}},
  \bibinfo{author}{\bibfnamefont{A.~E.} \bibnamefont{Roitberg}},
  \bibnamefont{and} \bibinfo{author}{\bibfnamefont{T.~P.}
  \bibnamefont{Martin}}, \bibinfo{journal}{J. Chem. Phys.}
  \textbf{\bibinfo{volume}{115}}, \bibinfo{pages}{3640} (\bibinfo{year}{2001}).

\bibitem[{\citenamefont{Doye and Hendy}(2003)}]{Doye03a}
\bibinfo{author}{\bibfnamefont{J.~P.~K.} \bibnamefont{Doye}} \bibnamefont{and}
  \bibinfo{author}{\bibfnamefont{S.~C.} \bibnamefont{Hendy}},
  \bibinfo{journal}{Eur. Phys. J. D} \textbf{\bibinfo{volume}{22}},
  \bibinfo{pages}{99} (\bibinfo{year}{2003}).

\bibitem[{\citenamefont{Garz\'{o}n et~al.}(1998)\citenamefont{Garz\'{o}n,
  Michaelian, and Beltr\'{a}n}}]{Garzon98}
\bibinfo{author}{\bibfnamefont{I.~L.} \bibnamefont{Garz\'{o}n}},
  \bibinfo{author}{\bibfnamefont{K.}~\bibnamefont{Michaelian}},
  \bibnamefont{and} \bibinfo{author}{\bibfnamefont{M.~R.}
  \bibnamefont{Beltr\'{a}n}}, \bibinfo{journal}{Phys. Rev. Lett.}
  \textbf{\bibinfo{volume}{81}}, \bibinfo{pages}{1600} (\bibinfo{year}{1998}).

\bibitem[{\citenamefont{Soler et~al.}(2001)\citenamefont{Soler, Garz\'{o}n, and
  Joannopoulos}}]{Soler01}
\bibinfo{author}{\bibfnamefont{J.~M.} \bibnamefont{Soler}},
  \bibinfo{author}{\bibfnamefont{I.~L.} \bibnamefont{Garz\'{o}n}},
  \bibnamefont{and} \bibinfo{author}{\bibfnamefont{J.~D.}
  \bibnamefont{Joannopoulos}}, \bibinfo{journal}{Solid State Commun.}
  \textbf{\bibinfo{volume}{117}}, \bibinfo{pages}{621} (\bibinfo{year}{2001}).

\bibitem[{\citenamefont{Lai et~al.}(2002)\citenamefont{Lai, Hsu, Wu, Liu, and
  Iwamatsu}}]{Lai02}
\bibinfo{author}{\bibfnamefont{S.~K.} \bibnamefont{Lai}},
  \bibinfo{author}{\bibfnamefont{P.~J.} \bibnamefont{Hsu}},
  \bibinfo{author}{\bibfnamefont{K.~L.} \bibnamefont{Wu}},
  \bibinfo{author}{\bibfnamefont{W.~K.} \bibnamefont{Liu}}, \bibnamefont{and}
  \bibinfo{author}{\bibfnamefont{M.}~\bibnamefont{Iwamatsu}},
  \bibinfo{journal}{J. Chem. Phys.} \textbf{\bibinfo{volume}{117}},
  \bibinfo{pages}{10715} (\bibinfo{year}{2002}).

\bibitem[{\citenamefont{Wilson and Johnston}(2000)}]{Wilson00}
\bibinfo{author}{\bibfnamefont{N.~T.} \bibnamefont{Wilson}} \bibnamefont{and}
  \bibinfo{author}{\bibfnamefont{R.~L.} \bibnamefont{Johnston}},
  \bibinfo{journal}{Eur. Phys. J. D} \textbf{\bibinfo{volume}{12}},
  \bibinfo{pages}{161} (\bibinfo{year}{2000}).

\bibitem[{\citenamefont{Doye and Wales}(1998)}]{Doye98c}
\bibinfo{author}{\bibfnamefont{J.~P.~K.} \bibnamefont{Doye}} \bibnamefont{and}
  \bibinfo{author}{\bibfnamefont{D.~J.} \bibnamefont{Wales}},
  \bibinfo{journal}{New J. Chem.} \textbf{\bibinfo{volume}{22}},
  \bibinfo{pages}{733} (\bibinfo{year}{1998}).

\bibitem[{\citenamefont{Michaelian et~al.}(2002)\citenamefont{Michaelian,
  Beltr\'{a}n, and Garz\'{o}n}}]{Michaelian02}
\bibinfo{author}{\bibfnamefont{K.}~\bibnamefont{Michaelian}},
  \bibinfo{author}{\bibfnamefont{M.~R.} \bibnamefont{Beltr\'{a}n}},
  \bibnamefont{and} \bibinfo{author}{\bibfnamefont{I.~L.}
  \bibnamefont{Garz\'{o}n}}, \bibinfo{journal}{Phys. Rev. B}
  \textbf{\bibinfo{volume}{65}}, \bibinfo{pages}{041403(R)}
  (\bibinfo{year}{2002}).

\bibitem[{\citenamefont{Soler et~al.}(2000)\citenamefont{Soler, Beltr\'{a}n,
  Michaelian, Garz\'{o}n, Ordej\'{o}n, S\'{a}nchez-Portal, and
  Artacho}}]{Soler00}
\bibinfo{author}{\bibfnamefont{J.~M.} \bibnamefont{Soler}},
  \bibinfo{author}{\bibfnamefont{M.~R.} \bibnamefont{Beltr\'{a}n}},
  \bibinfo{author}{\bibfnamefont{K.}~\bibnamefont{Michaelian}},
  \bibinfo{author}{\bibfnamefont{I.~L.} \bibnamefont{Garz\'{o}n}},
  \bibinfo{author}{\bibfnamefont{P.}~\bibnamefont{Ordej\'{o}n}},
  \bibinfo{author}{\bibfnamefont{D.}~\bibnamefont{S\'{a}nchez-Portal}},
  \bibnamefont{and} \bibinfo{author}{\bibfnamefont{E.}~\bibnamefont{Artacho}},
  \bibinfo{journal}{Phys. Rev. B} \textbf{\bibinfo{volume}{61}},
  \bibinfo{pages}{5771} (\bibinfo{year}{2000}).

\bibitem[{\citenamefont{Baletto et~al.}(2002)\citenamefont{Baletto, Ferrando,
  Fortunelli, Montalenti, and Mottet}}]{Baletto02b}
\bibinfo{author}{\bibfnamefont{F.}~\bibnamefont{Baletto}},
  \bibinfo{author}{\bibfnamefont{R.}~\bibnamefont{Ferrando}},
  \bibinfo{author}{\bibfnamefont{A.}~\bibnamefont{Fortunelli}},
  \bibinfo{author}{\bibfnamefont{F.}~\bibnamefont{Montalenti}},
  \bibnamefont{and} \bibinfo{author}{\bibfnamefont{C.}~\bibnamefont{Mottet}},
  \bibinfo{journal}{J. Chem. Phys.} \textbf{\bibinfo{volume}{116}},
  \bibinfo{pages}{3856} (\bibinfo{year}{2002}).

\bibitem[{\citenamefont{Doye and Wales}(1997)}]{Doye97d}
\bibinfo{author}{\bibfnamefont{J.~P.~K.} \bibnamefont{Doye}} \bibnamefont{and}
  \bibinfo{author}{\bibfnamefont{D.~J.} \bibnamefont{Wales}},
  \bibinfo{journal}{J. Chem. Soc., Faraday Trans.}
  \textbf{\bibinfo{volume}{93}}, \bibinfo{pages}{4233} (\bibinfo{year}{1997}).

\bibitem[{\citenamefont{Doye et~al.}(1995)\citenamefont{Doye, Wales, and
  Berry}}]{Doye95c}
\bibinfo{author}{\bibfnamefont{J.~P.~K.} \bibnamefont{Doye}},
  \bibinfo{author}{\bibfnamefont{D.~J.} \bibnamefont{Wales}}, \bibnamefont{and}
  \bibinfo{author}{\bibfnamefont{R.~S.} \bibnamefont{Berry}},
  \bibinfo{journal}{J. Chem. Phys.} \textbf{\bibinfo{volume}{103}},
  \bibinfo{pages}{4234} (\bibinfo{year}{1995}).

\bibitem[{\citenamefont{Doye and Wales}(2001)}]{Doye01d}
\bibinfo{author}{\bibfnamefont{J.~P.~K.} \bibnamefont{Doye}} \bibnamefont{and}
  \bibinfo{author}{\bibfnamefont{D.~J.} \bibnamefont{Wales}},
  \bibinfo{journal}{Phys. Rev. Lett.} \textbf{\bibinfo{volume}{86}},
  \bibinfo{pages}{5719} (\bibinfo{year}{2001}).

\bibitem[{\citenamefont{Doye et~al.}(2001)\citenamefont{Doye, Wales, and
  Simdyankin}}]{Doye01a}
\bibinfo{author}{\bibfnamefont{J.~P.~K.} \bibnamefont{Doye}},
  \bibinfo{author}{\bibfnamefont{D.~J.} \bibnamefont{Wales}}, \bibnamefont{and}
  \bibinfo{author}{\bibfnamefont{S.~I.} \bibnamefont{Simdyankin}},
  \bibinfo{journal}{Faraday Discuss.} \textbf{\bibinfo{volume}{118}},
  \bibinfo{pages}{159} (\bibinfo{year}{2001}).

\bibitem[{\citenamefont{Doye et~al.}(cond-mat/0205374)\citenamefont{Doye,
  Wales, Zetterling, and Dzugutov}}]{Doye03b}
\bibinfo{author}{\bibfnamefont{J.~P.~K.} \bibnamefont{Doye}},
  \bibinfo{author}{\bibfnamefont{D.~J.} \bibnamefont{Wales}},
  \bibinfo{author}{\bibfnamefont{F.~H.} \bibnamefont{Zetterling}},
  \bibnamefont{and} \bibinfo{author}{\bibfnamefont{M.}~\bibnamefont{Dzugutov}},
  \bibinfo{journal}{J. Chem. Phys.} p. \bibinfo{pages}{in press}
  (\bibinfo{year}{cond-mat/0205374}).

\bibitem[{\citenamefont{Dassenoy et~al.}(2000)\citenamefont{Dassenoy, Casanove,
  Lecante, Verelst, Snoeck, Mosset, Ould~Ely, Amiens, and
  Chaudret}}]{Dassenoy00}
\bibinfo{author}{\bibfnamefont{F.}~\bibnamefont{Dassenoy}},
  \bibinfo{author}{\bibfnamefont{M.-J.} \bibnamefont{Casanove}},
  \bibinfo{author}{\bibfnamefont{P.}~\bibnamefont{Lecante}},
  \bibinfo{author}{\bibfnamefont{M.}~\bibnamefont{Verelst}},
  \bibinfo{author}{\bibfnamefont{E.}~\bibnamefont{Snoeck}},
  \bibinfo{author}{\bibfnamefont{A.}~\bibnamefont{Mosset}},
  \bibinfo{author}{\bibfnamefont{T.}~\bibnamefont{Ould~Ely}},
  \bibinfo{author}{\bibfnamefont{C.}~\bibnamefont{Amiens}}, \bibnamefont{and}
  \bibinfo{author}{\bibfnamefont{B.}~\bibnamefont{Chaudret}},
  \bibinfo{journal}{J. Chem. Phys.} \textbf{\bibinfo{volume}{112}},
  \bibinfo{pages}{8137} (\bibinfo{year}{2000}).

\bibitem[{\citenamefont{Nelson and Spaepen}(1989)}]{NelsonS}
\bibinfo{author}{\bibfnamefont{D.~R.} \bibnamefont{Nelson}} \bibnamefont{and}
  \bibinfo{author}{\bibfnamefont{F.}~\bibnamefont{Spaepen}},
  \bibinfo{journal}{Solid State Phys.} \textbf{\bibinfo{volume}{42}},
  \bibinfo{pages}{1} (\bibinfo{year}{1989}).

\bibitem[{\citenamefont{Sadoc and Mosseri}(1999)}]{Sadoc99}
\bibinfo{author}{\bibfnamefont{J.-F.} \bibnamefont{Sadoc}} \bibnamefont{and}
  \bibinfo{author}{\bibfnamefont{R.}~\bibnamefont{Mosseri}},
  \emph{\bibinfo{title}{Geometric Frustration}} (\bibinfo{publisher}{Cambridge
  University Press}, \bibinfo{address}{Cambridge}, \bibinfo{year}{1999}).

\bibitem[{clo()}]{closep}
\bibinfo{note}{By contrast close-packed structures are composed of octahedra
  and tetrahedra.}

\bibitem[{\citenamefont{Br\'{e}chignac
  et~al.}(2000)\citenamefont{Br\'{e}chignac, Cahuzac, K\'{e}ba\"{i}li,
  Leygnier, and Yoshida}}]{Brechignac00}
\bibinfo{author}{\bibfnamefont{C.}~\bibnamefont{Br\'{e}chignac}},
  \bibinfo{author}{\bibfnamefont{P.}~\bibnamefont{Cahuzac}},
  \bibinfo{author}{\bibfnamefont{N.}~\bibnamefont{K\'{e}ba\"{i}li}},
  \bibinfo{author}{\bibfnamefont{J.}~\bibnamefont{Leygnier}}, \bibnamefont{and}
  \bibinfo{author}{\bibfnamefont{H.}~\bibnamefont{Yoshida}},
  \bibinfo{journal}{Phys. Rev. B} \textbf{\bibinfo{volume}{61}},
  \bibinfo{pages}{7280} (\bibinfo{year}{2000}).

\bibitem[{\citenamefont{Br\'{e}chignac
  et~al.}(1993)\citenamefont{Br\'{e}chignac, Cahuzac, Carlier, and
  Roux}}]{Brechignac93}
\bibinfo{author}{\bibfnamefont{C.}~\bibnamefont{Br\'{e}chignac}},
  \bibinfo{author}{\bibfnamefont{P.}~\bibnamefont{Cahuzac}},
  \bibinfo{author}{\bibfnamefont{F.}~\bibnamefont{Carlier}}, \bibnamefont{and}
  \bibinfo{author}{\bibfnamefont{J.~P.} \bibnamefont{Roux}},
  \bibinfo{journal}{Z. Phys. D} \textbf{\bibinfo{volume}{28}},
  \bibinfo{pages}{67} (\bibinfo{year}{1993}).

\bibitem[{\citenamefont{Shoemaker and Shoemaker}(1988)}]{Shoemaker}
\bibinfo{author}{\bibfnamefont{D.~P.} \bibnamefont{Shoemaker}}
  \bibnamefont{and} \bibinfo{author}{\bibfnamefont{C.~B.}
  \bibnamefont{Shoemaker}}, in \emph{\bibinfo{booktitle}{Introduction to
  Quasicrystals}}, edited by \bibinfo{editor}{\bibfnamefont{M.~V.}
  \bibnamefont{Jaric}} (\bibinfo{publisher}{Academic Press},
  \bibinfo{address}{London}, \bibinfo{year}{1988}), pp. \bibinfo{pages}{1--57}.

\bibitem[{bin()}]{binary}
\bibinfo{note}{The only exception is $\beta$-tungsten / $\beta$-W, which has
  the structure of the A15 Frank-Kasper phase.\cite{Shoemaker} However, two
  different valence states of the tungsten atoms are involved, giving it an
  effective binary character.}

\bibitem[{\citenamefont{Frank and Kasper}(1958)}]{FrankK58}
\bibinfo{author}{\bibfnamefont{F.~C.} \bibnamefont{Frank}} \bibnamefont{and}
  \bibinfo{author}{\bibfnamefont{J.~S.} \bibnamefont{Kasper}},
  \bibinfo{journal}{Acta Crystallogr.} \textbf{\bibinfo{volume}{11}},
  \bibinfo{pages}{184} (\bibinfo{year}{1958}).

\bibitem[{\citenamefont{Frank and Kasper}(1959)}]{FrankK59}
\bibinfo{author}{\bibfnamefont{F.~C.} \bibnamefont{Frank}} \bibnamefont{and}
  \bibinfo{author}{\bibfnamefont{J.~S.} \bibnamefont{Kasper}},
  \bibinfo{journal}{Acta Crystallogr.} \textbf{\bibinfo{volume}{12}},
  \bibinfo{pages}{483} (\bibinfo{year}{1959}).

\bibitem[{\citenamefont{Shechtman et~al.}(1984)\citenamefont{Shechtman, Blech,
  Gratias, and Cahn}}]{Shechtman84}
\bibinfo{author}{\bibfnamefont{D.}~\bibnamefont{Shechtman}},
  \bibinfo{author}{\bibfnamefont{I.}~\bibnamefont{Blech}},
  \bibinfo{author}{\bibfnamefont{D.}~\bibnamefont{Gratias}}, \bibnamefont{and}
  \bibinfo{author}{\bibfnamefont{J.~W.} \bibnamefont{Cahn}},
  \bibinfo{journal}{Phys. Rev. Lett.} \textbf{\bibinfo{volume}{53}},
  \bibinfo{pages}{1951} (\bibinfo{year}{1984}).

\bibitem[{\citenamefont{Elser and Henley}(1985)}]{Elser85}
\bibinfo{author}{\bibfnamefont{V.}~\bibnamefont{Elser}} \bibnamefont{and}
  \bibinfo{author}{\bibfnamefont{C.~L.} \bibnamefont{Henley}},
  \bibinfo{journal}{Phys. Rev. Lett.} \textbf{\bibinfo{volume}{55}},
  \bibinfo{pages}{2883} (\bibinfo{year}{1985}).

\bibitem[{\citenamefont{Sachdev and Nelson}(1985)}]{Sachdev85b}
\bibinfo{author}{\bibfnamefont{S.}~\bibnamefont{Sachdev}} \bibnamefont{and}
  \bibinfo{author}{\bibfnamefont{D.~R.} \bibnamefont{Nelson}},
  \bibinfo{journal}{Phys. Rev. B} \textbf{\bibinfo{volume}{32}},
  \bibinfo{pages}{4592} (\bibinfo{year}{1985}).

\bibitem[{\citenamefont{Audier and Guyot}(1986)}]{Audier86}
\bibinfo{author}{\bibfnamefont{M.}~\bibnamefont{Audier}} \bibnamefont{and}
  \bibinfo{author}{\bibfnamefont{P.}~\bibnamefont{Guyot}},
  \bibinfo{journal}{Philos. Mag. B} \textbf{\bibinfo{volume}{53}},
  \bibinfo{pages}{L43} (\bibinfo{year}{1986}).

\bibitem[{\citenamefont{Northby}(1987)}]{Northby87}
\bibinfo{author}{\bibfnamefont{J.~A.} \bibnamefont{Northby}},
  \bibinfo{journal}{J. Chem. Phys.} \textbf{\bibinfo{volume}{87}},
  \bibinfo{pages}{6166} (\bibinfo{year}{1987}).

\bibitem[{dis()}]{disclin}
\bibinfo{note}{Positive disclinations run along edges that are common to only
  four tetrahedra. As this type of disclination does not occur in any of the
  polytetrahedral structures that I consider, in the rest of this paper the
  term disclination is taken to mean a negative disclination.}

\bibitem[{maj()}]{major}
\bibinfo{note}{In Frank-Kasper phases the disclination network is sometimes
  called the major network.}

\bibitem[{\citenamefont{Hafner}(1998)}]{Hafner88b}
\bibinfo{author}{\bibfnamefont{J.}~\bibnamefont{Hafner}},
  \emph{\bibinfo{title}{From Hamiltonians to phase diagrams}}
  (\bibinfo{publisher}{Springer-Verlag}, \bibinfo{address}{Berlin},
  \bibinfo{year}{1998}).

\bibitem[{\citenamefont{Moriarty and Widom}(1997)}]{Moriarty97}
\bibinfo{author}{\bibfnamefont{J.~A.} \bibnamefont{Moriarty}} \bibnamefont{and}
  \bibinfo{author}{\bibfnamefont{M.}~\bibnamefont{Widom}},
  \bibinfo{journal}{Phys. Rev. B} \textbf{\bibinfo{volume}{56}},
  \bibinfo{pages}{7905} (\bibinfo{year}{1997}).

\bibitem[{\citenamefont{Cune and Apostol}(2000)}]{Cune00}
\bibinfo{author}{\bibfnamefont{L.~C.} \bibnamefont{Cune}} \bibnamefont{and}
  \bibinfo{author}{\bibfnamefont{M.}~\bibnamefont{Apostol}},
  \bibinfo{journal}{Phys. Lett. A} \textbf{\bibinfo{volume}{273}},
  \bibinfo{pages}{117} (\bibinfo{year}{2000}).

\bibitem[{\citenamefont{Ercolessi and Adams}(1994)}]{Ercolessi94}
\bibinfo{author}{\bibfnamefont{F.}~\bibnamefont{Ercolessi}} \bibnamefont{and}
  \bibinfo{author}{\bibfnamefont{J.~B.} \bibnamefont{Adams}},
  \bibinfo{journal}{Europhys. Lett.} \textbf{\bibinfo{volume}{26}},
  \bibinfo{pages}{583} (\bibinfo{year}{1994}).

\bibitem[{\citenamefont{Ercolessi et~al.}(1988)\citenamefont{Ercolessi,
  Parrinello, and Tosatti}}]{Ercolessi88}
\bibinfo{author}{\bibfnamefont{F.}~\bibnamefont{Ercolessi}},
  \bibinfo{author}{\bibfnamefont{M.}~\bibnamefont{Parrinello}},
  \bibnamefont{and} \bibinfo{author}{\bibfnamefont{E.}~\bibnamefont{Tosatti}},
  \bibinfo{journal}{Philos. Mag. A} \textbf{\bibinfo{volume}{58}},
  \bibinfo{pages}{213} (\bibinfo{year}{1988}).

\bibitem[{\citenamefont{Sun and Gong}(1993)}]{Sun98}
\bibinfo{author}{\bibfnamefont{D.~Y.} \bibnamefont{Sun}} \bibnamefont{and}
  \bibinfo{author}{\bibfnamefont{X.~G.} \bibnamefont{Gong}},
  \bibinfo{journal}{Phys. Rev. B} \textbf{\bibinfo{volume}{98}},
  \bibinfo{pages}{9707} (\bibinfo{year}{1993}).

\bibitem[{\citenamefont{G\"{u}lseren et~al.}(1998)\citenamefont{G\"{u}lseren,
  Ercolessi, and Tosatti}}]{Gulseren98}
\bibinfo{author}{\bibfnamefont{O.}~\bibnamefont{G\"{u}lseren}},
  \bibinfo{author}{\bibfnamefont{F.}~\bibnamefont{Ercolessi}},
  \bibnamefont{and} \bibinfo{author}{\bibfnamefont{E.}~\bibnamefont{Tosatti}},
  \bibinfo{journal}{Phys. Rev. Lett.} \textbf{\bibinfo{volume}{80}},
  \bibinfo{pages}{3775} (\bibinfo{year}{1998}).

\bibitem[{\citenamefont{Di~Tolla et~al.}(1995)\citenamefont{Di~Tolla,
  Ercolessi, and Tosatti}}]{DiTolla94}
\bibinfo{author}{\bibfnamefont{F.~D.} \bibnamefont{Di~Tolla}},
  \bibinfo{author}{\bibfnamefont{F.}~\bibnamefont{Ercolessi}},
  \bibnamefont{and} \bibinfo{author}{\bibfnamefont{E.}~\bibnamefont{Tosatti}},
  \bibinfo{journal}{Phys. Rev. Lett.} \textbf{\bibinfo{volume}{74}},
  \bibinfo{pages}{3201} (\bibinfo{year}{1995}).

\bibitem[{\citenamefont{Raphuthi et~al.}(1995)\citenamefont{Raphuthi, Wang,
  Ercolessi, and Adams}}]{Raphuthi95}
\bibinfo{author}{\bibfnamefont{A.~M.} \bibnamefont{Raphuthi}},
  \bibinfo{author}{\bibfnamefont{X.~Q.} \bibnamefont{Wang}},
  \bibinfo{author}{\bibfnamefont{F.}~\bibnamefont{Ercolessi}},
  \bibnamefont{and} \bibinfo{author}{\bibfnamefont{J.~B.} \bibnamefont{Adams}},
  \bibinfo{journal}{Phys. Rev. B} \textbf{\bibinfo{volume}{52}},
  \bibinfo{pages}{R5554} (\bibinfo{year}{1995}).

\bibitem[{\citenamefont{Rittner et~al.}(1996)\citenamefont{Rittner, Seidman,
  and Merkle}}]{Rittner96}
\bibinfo{author}{\bibfnamefont{J.~D.} \bibnamefont{Rittner}},
  \bibinfo{author}{\bibfnamefont{D.~N.} \bibnamefont{Seidman}},
  \bibnamefont{and} \bibinfo{author}{\bibfnamefont{K.~L.}
  \bibnamefont{Merkle}}, \bibinfo{journal}{Phys. Rev. B}
  \textbf{\bibinfo{volume}{53}}, \bibinfo{pages}{R4241} (\bibinfo{year}{1996}).

\bibitem[{\citenamefont{Sandberg et~al.}(2002)\citenamefont{Sandberg,
  Magyari-K\"{o}pe, and Mattson}}]{Sandberg02}
\bibinfo{author}{\bibfnamefont{N.}~\bibnamefont{Sandberg}},
  \bibinfo{author}{\bibfnamefont{B.}~\bibnamefont{Magyari-K\"{o}pe}},
  \bibnamefont{and} \bibinfo{author}{\bibfnamefont{T.~R.}
  \bibnamefont{Mattson}}, \bibinfo{journal}{Phys. Rev. Lett}
  \textbf{\bibinfo{volume}{89}}, \bibinfo{pages}{065901}
  (\bibinfo{year}{2002}).

\bibitem[{\citenamefont{Trushin et~al.}(2001)\citenamefont{Trushin, Salo,
  Alatalo, and Ala-Nissila}}]{Trushin01}
\bibinfo{author}{\bibfnamefont{O.~S.} \bibnamefont{Trushin}},
  \bibinfo{author}{\bibfnamefont{P.}~\bibnamefont{Salo}},
  \bibinfo{author}{\bibfnamefont{M.}~\bibnamefont{Alatalo}}, \bibnamefont{and}
  \bibinfo{author}{\bibfnamefont{T.}~\bibnamefont{Ala-Nissila}},
  \bibinfo{journal}{Surf. Sci.} \textbf{\bibinfo{volume}{482}},
  \bibinfo{pages}{365} (\bibinfo{year}{2001}).

\bibitem[{\citenamefont{Liu et~al.}(2001)\citenamefont{Liu, Zhu, Xia, and
  Sun}}]{Liu01}
\bibinfo{author}{\bibfnamefont{C.~S.} \bibnamefont{Liu}},
  \bibinfo{author}{\bibfnamefont{Z.~G.} \bibnamefont{Zhu}},
  \bibinfo{author}{\bibfnamefont{J.}~\bibnamefont{Xia}}, \bibnamefont{and}
  \bibinfo{author}{\bibfnamefont{D.~Y.} \bibnamefont{Sun}},
  \bibinfo{journal}{J. Phys.: Condens. Matter} \textbf{\bibinfo{volume}{13}},
  \bibinfo{pages}{1873} (\bibinfo{year}{2001}).

\bibitem[{\citenamefont{Mishin et~al.}(1999)\citenamefont{Mishin, Farkas, Mehl,
  and Papaconstantopoulos}}]{Mishin99}
\bibinfo{author}{\bibfnamefont{Y.}~\bibnamefont{Mishin}},
  \bibinfo{author}{\bibfnamefont{D.}~\bibnamefont{Farkas}},
  \bibinfo{author}{\bibfnamefont{M.~J.} \bibnamefont{Mehl}}, \bibnamefont{and}
  \bibinfo{author}{\bibfnamefont{D.~A.} \bibnamefont{Papaconstantopoulos}},
  \bibinfo{journal}{Phys. Rev. B} \textbf{\bibinfo{volume}{59}},
  \bibinfo{pages}{3393} (\bibinfo{year}{1999}).

\bibitem[{\citenamefont{Lim et~al.}(1992)\citenamefont{Lim, Ong, and
  Ercolessi}}]{Lim}
\bibinfo{author}{\bibfnamefont{H.~S.} \bibnamefont{Lim}},
  \bibinfo{author}{\bibfnamefont{C.~K.} \bibnamefont{Ong}}, \bibnamefont{and}
  \bibinfo{author}{\bibfnamefont{F.}~\bibnamefont{Ercolessi}},
  \bibinfo{journal}{Surf. Sci.} \textbf{\bibinfo{volume}{269/270}},
  \bibinfo{pages}{1109} (\bibinfo{year}{1992}).

\bibitem[{\citenamefont{Hendy and Doye}(2002)}]{Hendy02}
\bibinfo{author}{\bibfnamefont{S.~C.} \bibnamefont{Hendy}} \bibnamefont{and}
  \bibinfo{author}{\bibfnamefont{J.~P.~K.} \bibnamefont{Doye}},
  \bibinfo{journal}{Phys. Rev. B} \textbf{\bibinfo{volume}{66}},
  \bibinfo{pages}{235402} (\bibinfo{year}{2002}).

\bibitem[{\citenamefont{Wales and Doye}(1997)}]{WalesD97}
\bibinfo{author}{\bibfnamefont{D.~J.} \bibnamefont{Wales}} \bibnamefont{and}
  \bibinfo{author}{\bibfnamefont{J.~P.~K.} \bibnamefont{Doye}},
  \bibinfo{journal}{J. Phys. Chem. A} \textbf{\bibinfo{volume}{101}},
  \bibinfo{pages}{5111} (\bibinfo{year}{1997}).

\bibitem[{\citenamefont{Li and Scheraga}(1987)}]{Li87a}
\bibinfo{author}{\bibfnamefont{Z.}~\bibnamefont{Li}} \bibnamefont{and}
  \bibinfo{author}{\bibfnamefont{H.~A.} \bibnamefont{Scheraga}},
  \bibinfo{journal}{Proc. Natl. Acad. Sci. USA} \textbf{\bibinfo{volume}{84}},
  \bibinfo{pages}{6611} (\bibinfo{year}{1987}).

\bibitem[{\citenamefont{Tsai and Jordan}(1993)}]{Tsai93a}
\bibinfo{author}{\bibfnamefont{C.~J.} \bibnamefont{Tsai}} \bibnamefont{and}
  \bibinfo{author}{\bibfnamefont{K.~D.} \bibnamefont{Jordan}},
  \bibinfo{journal}{J. Phys. Chem.} \textbf{\bibinfo{volume}{97}},
  \bibinfo{pages}{11227} (\bibinfo{year}{1993}).

\bibitem[{\citenamefont{Stillinger}(1999)}]{Still99}
\bibinfo{author}{\bibfnamefont{F.~H.} \bibnamefont{Stillinger}},
  \bibinfo{journal}{Phys. Rev. E} \textbf{\bibinfo{volume}{59}},
  \bibinfo{pages}{48} (\bibinfo{year}{1999}).

\bibitem[{\citenamefont{Doye and Wales}(2002{\natexlab{a}})}]{Doye02a}
\bibinfo{author}{\bibfnamefont{J.~P.~K.} \bibnamefont{Doye}} \bibnamefont{and}
  \bibinfo{author}{\bibfnamefont{D.~J.} \bibnamefont{Wales}},
  \bibinfo{journal}{J. Chem. Phys.} \textbf{\bibinfo{volume}{116}},
  \bibinfo{pages}{3777} (\bibinfo{year}{2002}{\natexlab{a}}).

\bibitem[{Web()}]{Web}
\bibinfo{note}{D. J. Wales, J. P. K. Doye, A. Dullweber, M. P. Hodges, F. Y.
  Naumkin, F. Calvo, J. Hern\'{a}ndez-Rojas and T. F. Middleton, The Cambridge
  Cluster Database, http://www-wales.ch.cam.ac.uk/CCD.html}.

\bibitem[{\citenamefont{Stampfli}(1986)}]{Stampfli88}
\bibinfo{author}{\bibfnamefont{P.}~\bibnamefont{Stampfli}},
  \bibinfo{journal}{Helv. Phys. Acta} \textbf{\bibinfo{volume}{59}},
  \bibinfo{pages}{1260} (\bibinfo{year}{1986}).

\bibitem[{\citenamefont{Dzugutov}(1993)}]{Dzugutov93}
\bibinfo{author}{\bibfnamefont{M.}~\bibnamefont{Dzugutov}},
  \bibinfo{journal}{Phys. Rev. Lett.} \textbf{\bibinfo{volume}{70}},
  \bibinfo{pages}{2924} (\bibinfo{year}{1993}).

\bibitem[{\citenamefont{Roth and G\"{ahler}}(1998)}]{Roth98}
\bibinfo{author}{\bibfnamefont{J.}~\bibnamefont{Roth}} \bibnamefont{and}
  \bibinfo{author}{\bibfnamefont{F.}~\bibnamefont{G\"{ahler}}},
  \bibinfo{journal}{Eur. Phys. J. B} \textbf{\bibinfo{volume}{6}},
  \bibinfo{pages}{425} (\bibinfo{year}{1998}).

\bibitem[{str()}]{strain}
\bibinfo{note}{A similar decomposition has been used for pair
  potentials,\cite{Doye95c,Doye97d} except that the correction term, $E_{\rm
  strain}$, is with respect to the energy that would result if all
  nearest-neighbour pairs were at the equilibrium pair separation, not the
  average value.}

\bibitem[{\citenamefont{Raoult et~al.}(1989)\citenamefont{Raoult, Farges,
  de~Feraudy, and Torchet}}]{Raoult89a}
\bibinfo{author}{\bibfnamefont{B.}~\bibnamefont{Raoult}},
  \bibinfo{author}{\bibfnamefont{J.}~\bibnamefont{Farges}},
  \bibinfo{author}{\bibfnamefont{M.-F.} \bibnamefont{de~Feraudy}},
  \bibnamefont{and} \bibinfo{author}{\bibfnamefont{G.}~\bibnamefont{Torchet}},
  \bibinfo{journal}{Philos. Mag. B} \textbf{\bibinfo{volume}{60}},
  \bibinfo{pages}{881} (\bibinfo{year}{1989}).

\bibitem[{\citenamefont{Schriver et~al.}(1990)\citenamefont{Schriver, Persson,
  Honea, and Whetten}}]{Schriver90}
\bibinfo{author}{\bibfnamefont{K.~E.} \bibnamefont{Schriver}},
  \bibinfo{author}{\bibfnamefont{J.~L.} \bibnamefont{Persson}},
  \bibinfo{author}{\bibfnamefont{E.~C.} \bibnamefont{Honea}}, \bibnamefont{and}
  \bibinfo{author}{\bibfnamefont{R.~L.} \bibnamefont{Whetten}},
  \bibinfo{journal}{Phys. Rev. Lett} \textbf{\bibinfo{volume}{64}},
  \bibinfo{pages}{2539} (\bibinfo{year}{1990}).

\bibitem[{\citenamefont{Pellarin et~al.}(1993)\citenamefont{Pellarin,
  Baguenard, Broyer, Lerm\'e, Vialle, and Perez}}]{Pellarin93}
\bibinfo{author}{\bibfnamefont{M.}~\bibnamefont{Pellarin}},
  \bibinfo{author}{\bibfnamefont{B.}~\bibnamefont{Baguenard}},
  \bibinfo{author}{\bibfnamefont{M.}~\bibnamefont{Broyer}},
  \bibinfo{author}{\bibfnamefont{J.}~\bibnamefont{Lerm\'e}},
  \bibinfo{author}{\bibfnamefont{J.~L.} \bibnamefont{Vialle}},
  \bibnamefont{and} \bibinfo{author}{\bibfnamefont{A.}~\bibnamefont{Perez}},
  \bibinfo{journal}{J. Chem. Phys.} \textbf{\bibinfo{volume}{98}},
  \bibinfo{pages}{944} (\bibinfo{year}{1993}).

\bibitem[{\citenamefont{Jarrold and Bower}(1993)}]{Jarrold93}
\bibinfo{author}{\bibfnamefont{M.~F.} \bibnamefont{Jarrold}} \bibnamefont{and}
  \bibinfo{author}{\bibfnamefont{J.~E.} \bibnamefont{Bower}},
  \bibinfo{journal}{J. Chem. Phys.} \textbf{\bibinfo{volume}{98}},
  \bibinfo{pages}{2399} (\bibinfo{year}{1993}).

\bibitem[{\citenamefont{Akola et~al.}(2000)\citenamefont{Akola, Maninnen,
  H\"{a}kkinen, Landman, Li, and Wang}}]{Akola00}
\bibinfo{author}{\bibfnamefont{J.}~\bibnamefont{Akola}},
  \bibinfo{author}{\bibnamefont{Maninnen}},
  \bibinfo{author}{\bibfnamefont{H.}~\bibnamefont{H\"{a}kkinen}},
  \bibinfo{author}{\bibfnamefont{U.}~\bibnamefont{Landman}},
  \bibinfo{author}{\bibfnamefont{X.}~\bibnamefont{Li}}, \bibnamefont{and}
  \bibinfo{author}{\bibfnamefont{L.-S.} \bibnamefont{Wang}},
  \bibinfo{journal}{Phys. Rev. B} \textbf{\bibinfo{volume}{62}},
  \bibinfo{pages}{13216} (\bibinfo{year}{2000}).

\bibitem[{\citenamefont{Li et~al.}(1998)\citenamefont{Li, Wu, Wang, and
  Wang}}]{Li98}
\bibinfo{author}{\bibfnamefont{X.}~\bibnamefont{Li}},
  \bibinfo{author}{\bibfnamefont{H.}~\bibnamefont{Wu}},
  \bibinfo{author}{\bibfnamefont{X.-B.} \bibnamefont{Wang}}, \bibnamefont{and}
  \bibinfo{author}{\bibfnamefont{L.-S.} \bibnamefont{Wang}},
  \bibinfo{journal}{Phys. Rev. Lett.} \textbf{\bibinfo{volume}{81}},
  \bibinfo{pages}{1909} (\bibinfo{year}{1998}).

\bibitem[{\citenamefont{Akola et~al.}(1999)\citenamefont{Akola, Maninnen,
  H\"{a}kkinen, Landman, Li, and Wang}}]{Akola99}
\bibinfo{author}{\bibfnamefont{J.}~\bibnamefont{Akola}},
  \bibinfo{author}{\bibnamefont{Maninnen}},
  \bibinfo{author}{\bibfnamefont{H.}~\bibnamefont{H\"{a}kkinen}},
  \bibinfo{author}{\bibfnamefont{U.}~\bibnamefont{Landman}},
  \bibinfo{author}{\bibfnamefont{X.}~\bibnamefont{Li}}, \bibnamefont{and}
  \bibinfo{author}{\bibfnamefont{L.-S.} \bibnamefont{Wang}},
  \bibinfo{journal}{Phys. Rev. B} \textbf{\bibinfo{volume}{60}},
  \bibinfo{pages}{R11297} (\bibinfo{year}{1999}).

\bibitem[{\citenamefont{Lloyd et~al.}(2002)\citenamefont{Lloyd, Johnston,
  Roberts, and Mortimer-Jones}}]{Lloyd02}
\bibinfo{author}{\bibfnamefont{L.~D.} \bibnamefont{Lloyd}},
  \bibinfo{author}{\bibfnamefont{R.~L.} \bibnamefont{Johnston}},
  \bibinfo{author}{\bibfnamefont{C.}~\bibnamefont{Roberts}}, \bibnamefont{and}
  \bibinfo{author}{\bibfnamefont{T.~V.} \bibnamefont{Mortimer-Jones}},
  \bibinfo{journal}{CHEMPHYSCHEM} \textbf{\bibinfo{volume}{3}},
  \bibinfo{pages}{408} (\bibinfo{year}{2002}).

\bibitem[{\citenamefont{Lloyd and Johnston}(1998)}]{Lloyd98}
\bibinfo{author}{\bibfnamefont{L.~D.} \bibnamefont{Lloyd}} \bibnamefont{and}
  \bibinfo{author}{\bibfnamefont{R.~L.} \bibnamefont{Johnston}},
  \bibinfo{journal}{Chem. Phys.} \textbf{\bibinfo{volume}{236}},
  \bibinfo{pages}{107} (\bibinfo{year}{1998}).

\bibitem[{\citenamefont{Bailey et~al.}(2003)\citenamefont{Bailey, Wilson,
  Roberts, and Johnston}}]{Bailey03}
\bibinfo{author}{\bibfnamefont{M.~S.} \bibnamefont{Bailey}},
  \bibinfo{author}{\bibfnamefont{N.~T.} \bibnamefont{Wilson}},
  \bibinfo{author}{\bibfnamefont{C.}~\bibnamefont{Roberts}}, \bibnamefont{and}
  \bibinfo{author}{\bibfnamefont{R.~L.} \bibnamefont{Johnston}},
  \bibinfo{journal}{Eur. Phys. J. D}  (\bibinfo{year}{2003}).

\bibitem[{\citenamefont{Akola et~al.}(1998)\citenamefont{Akola, H\"{a}kkinen,
  and Maninnen}}]{Akola98}
\bibinfo{author}{\bibfnamefont{J.}~\bibnamefont{Akola}},
  \bibinfo{author}{\bibfnamefont{H.}~\bibnamefont{H\"{a}kkinen}},
  \bibnamefont{and} \bibinfo{author}{\bibnamefont{Maninnen}},
  \bibinfo{journal}{Phys. Rev. B} \textbf{\bibinfo{volume}{58}},
  \bibinfo{pages}{3601} (\bibinfo{year}{1998}).

\bibitem[{\citenamefont{Ahlrichs and Elliott}(1999)}]{Ahlrichs99}
\bibinfo{author}{\bibfnamefont{R.}~\bibnamefont{Ahlrichs}} \bibnamefont{and}
  \bibinfo{author}{\bibfnamefont{S.~D.} \bibnamefont{Elliott}},
  \bibinfo{journal}{Phys. Chem. Chem. Phys.} \textbf{\bibinfo{volume}{1}},
  \bibinfo{pages}{13} (\bibinfo{year}{1999}).

\bibitem[{\citenamefont{Cheng et~al.}(1991)\citenamefont{Cheng, Berry, and
  Whetten}}]{Cheng91}
\bibinfo{author}{\bibfnamefont{H.-P.} \bibnamefont{Cheng}},
  \bibinfo{author}{\bibfnamefont{R.~S.} \bibnamefont{Berry}}, \bibnamefont{and}
  \bibinfo{author}{\bibfnamefont{R.~L.} \bibnamefont{Whetten}},
  \bibinfo{journal}{Phys. Rev. B} \textbf{\bibinfo{volume}{43}},
  \bibinfo{pages}{10647} (\bibinfo{year}{1991}).

\bibitem[{\citenamefont{Jones}(1991)}]{Jones91}
\bibinfo{author}{\bibfnamefont{R.~O.} \bibnamefont{Jones}},
  \bibinfo{journal}{Phys. Rev. Lett.} \textbf{\bibinfo{volume}{67}},
  \bibinfo{pages}{224} (\bibinfo{year}{1991}).

\bibitem[{\citenamefont{Yi et~al.}(1991)\citenamefont{Yi, Oh, and
  Bernholc}}]{Yi91}
\bibinfo{author}{\bibfnamefont{J.-Y.} \bibnamefont{Yi}},
  \bibinfo{author}{\bibfnamefont{D.~J.} \bibnamefont{Oh}}, \bibnamefont{and}
  \bibinfo{author}{\bibfnamefont{J.}~\bibnamefont{Bernholc}},
  \bibinfo{journal}{Phys. Rev. Lett.} \textbf{\bibinfo{volume}{67}},
  \bibinfo{pages}{1594} (\bibinfo{year}{1991}).

\bibitem[{\citenamefont{Debiaggi and Caro}(1992)}]{Debiaggi92}
\bibinfo{author}{\bibfnamefont{S.}~\bibnamefont{Debiaggi}} \bibnamefont{and}
  \bibinfo{author}{\bibfnamefont{A.}~\bibnamefont{Caro}},
  \bibinfo{journal}{Phys. Rev. B} \textbf{\bibinfo{volume}{46}},
  \bibinfo{pages}{7322} (\bibinfo{year}{1992}).

\bibitem[{\citenamefont{Yang et~al.}(1993)\citenamefont{Yang, Drabold, Adams,
  and Sachdev}}]{Yang93}
\bibinfo{author}{\bibfnamefont{S.~H.} \bibnamefont{Yang}},
  \bibinfo{author}{\bibfnamefont{D.~A.} \bibnamefont{Drabold}},
  \bibinfo{author}{\bibfnamefont{J.~B.} \bibnamefont{Adams}}, \bibnamefont{and}
  \bibinfo{author}{\bibfnamefont{A.}~\bibnamefont{Sachdev}},
  \bibinfo{journal}{Phys. Rev. B} \textbf{\bibinfo{volume}{47}},
  \bibinfo{pages}{1567} (\bibinfo{year}{1993}).

\bibitem[{\citenamefont{Martin et~al.}(1992)\citenamefont{Martin, N\"aher, and
  Schaber}}]{Martin92}
\bibinfo{author}{\bibfnamefont{T.~P.} \bibnamefont{Martin}},
  \bibinfo{author}{\bibfnamefont{U.}~\bibnamefont{N\"aher}}, \bibnamefont{and}
  \bibinfo{author}{\bibfnamefont{H.}~\bibnamefont{Schaber}},
  \bibinfo{journal}{Chem. Phys. Lett.} \textbf{\bibinfo{volume}{199}},
  \bibinfo{pages}{470} (\bibinfo{year}{1992}).

\bibitem[{\citenamefont{Baguenard et~al.}(1994)\citenamefont{Baguenard,
  Pellarin, Lerm\'e, Vialle, and Broyer}}]{Baguenard}
\bibinfo{author}{\bibfnamefont{B.}~\bibnamefont{Baguenard}},
  \bibinfo{author}{\bibfnamefont{M.}~\bibnamefont{Pellarin}},
  \bibinfo{author}{\bibfnamefont{J.}~\bibnamefont{Lerm\'e}},
  \bibinfo{author}{\bibfnamefont{J.~L.} \bibnamefont{Vialle}},
  \bibnamefont{and} \bibinfo{author}{\bibfnamefont{M.}~\bibnamefont{Broyer}},
  \bibinfo{journal}{J. Chem. Phys.} \textbf{\bibinfo{volume}{100}},
  \bibinfo{pages}{754} (\bibinfo{year}{1994}).

\bibitem[{\citenamefont{Lerm\'e et~al.}(1992)\citenamefont{Lerm\'e, Pellarin,
  Vialle, Baguenard, and Broyer}}]{Lerme92}
\bibinfo{author}{\bibfnamefont{J.}~\bibnamefont{Lerm\'e}},
  \bibinfo{author}{\bibfnamefont{M.}~\bibnamefont{Pellarin}},
  \bibinfo{author}{\bibfnamefont{J.~L.} \bibnamefont{Vialle}},
  \bibinfo{author}{\bibfnamefont{B.}~\bibnamefont{Baguenard}},
  \bibnamefont{and} \bibinfo{author}{\bibfnamefont{M.}~\bibnamefont{Broyer}},
  \bibinfo{journal}{Phys. Rev. Lett.} \textbf{\bibinfo{volume}{68}},
  \bibinfo{pages}{2818} (\bibinfo{year}{1992}).

\bibitem[{\citenamefont{N\"{a}her et~al.}(1993)\citenamefont{N\"{a}her,
  Zimmermann, and Martin}}]{Naher93}
\bibinfo{author}{\bibfnamefont{U.}~\bibnamefont{N\"{a}her}},
  \bibinfo{author}{\bibfnamefont{U.}~\bibnamefont{Zimmermann}},
  \bibnamefont{and} \bibinfo{author}{\bibfnamefont{T.~P.}
  \bibnamefont{Martin}}, \bibinfo{journal}{J. Chem. Phys.}
  \textbf{\bibinfo{volume}{99}}, \bibinfo{pages}{2256} (\bibinfo{year}{1993}).

\bibitem[{\citenamefont{Valkealahti and Manninen}(1994)}]{Valkealahti94}
\bibinfo{author}{\bibfnamefont{S.}~\bibnamefont{Valkealahti}} \bibnamefont{and}
  \bibinfo{author}{\bibfnamefont{M.}~\bibnamefont{Manninen}},
  \bibinfo{journal}{Phys. Rev. B} \textbf{\bibinfo{volume}{50}},
  \bibinfo{pages}{17564} (\bibinfo{year}{1994}).

\bibitem[{\citenamefont{Valkealahti et~al.}(1995)\citenamefont{Valkealahti,
  N\"{a}her, and Manninen}}]{Valkealahti95}
\bibinfo{author}{\bibfnamefont{S.}~\bibnamefont{Valkealahti}},
  \bibinfo{author}{\bibfnamefont{U.}~\bibnamefont{N\"{a}her}},
  \bibnamefont{and} \bibinfo{author}{\bibfnamefont{M.}~\bibnamefont{Manninen}},
  \bibinfo{journal}{Phys. Rev. B} \textbf{\bibinfo{volume}{51}},
  \bibinfo{pages}{11039} (\bibinfo{year}{1995}).

\bibitem[{\citenamefont{Valkealahti and Manninen}(1998)}]{Valkealahti98}
\bibinfo{author}{\bibfnamefont{S.}~\bibnamefont{Valkealahti}} \bibnamefont{and}
  \bibinfo{author}{\bibfnamefont{M.}~\bibnamefont{Manninen}},
  \bibinfo{journal}{Phys. Rev. B} \textbf{\bibinfo{volume}{57}},
  \bibinfo{pages}{15533} (\bibinfo{year}{1998}).

\bibitem[{\citenamefont{van~de Waal}(1989)}]{van89}
\bibinfo{author}{\bibfnamefont{B.~W.} \bibnamefont{van~de Waal}},
  \bibinfo{journal}{J. Chem. Phys.} \textbf{\bibinfo{volume}{90}},
  \bibinfo{pages}{3407} (\bibinfo{year}{1989}).

\bibitem[{\citenamefont{Turner et~al.}(2000)\citenamefont{Turner, Johnston, and
  Wilson}}]{Turner00}
\bibinfo{author}{\bibfnamefont{G.~W.} \bibnamefont{Turner}},
  \bibinfo{author}{\bibfnamefont{R.~L.} \bibnamefont{Johnston}},
  \bibnamefont{and} \bibinfo{author}{\bibfnamefont{N.~T.}
  \bibnamefont{Wilson}}, \bibinfo{journal}{J. Chem. Phys.}
  \textbf{\bibinfo{volume}{112}}, \bibinfo{pages}{4773} (\bibinfo{year}{2000}).

\bibitem[{\citenamefont{Doye and Calvo}(2001)}]{Doye01b}
\bibinfo{author}{\bibfnamefont{J.~P.~K.} \bibnamefont{Doye}} \bibnamefont{and}
  \bibinfo{author}{\bibfnamefont{F.}~\bibnamefont{Calvo}},
  \bibinfo{journal}{Phys. Rev. Lett.} \textbf{\bibinfo{volume}{86}},
  \bibinfo{pages}{3570} (\bibinfo{year}{2001}).

\bibitem[{\citenamefont{Doye and Wales}(2002{\natexlab{b}})}]{Doye02b}
\bibinfo{author}{\bibfnamefont{J.~P.~K.} \bibnamefont{Doye}} \bibnamefont{and}
  \bibinfo{author}{\bibfnamefont{D.~J.} \bibnamefont{Wales}},
  \bibinfo{journal}{J. Chem. Phys.} \textbf{\bibinfo{volume}{116}},
  \bibinfo{pages}{8307} (\bibinfo{year}{2002}{\natexlab{b}}).

\end{thebibliography}
\end{document}